\documentclass[11pt]{article}
\usepackage{amsfonts,amssymb,amsmath
}

\hsize \textwidth
\advance \hsize by -\marginparwidth
\evensidemargin \oddsidemargin
\usepackage{amssymb}
\advance\hoffset by 5mm

\newtheorem{df}{Definition}[section]
\newtheorem{lm}[df]{Lemma}

\newtheorem{prop}[df]{Proposition}
\newtheorem{thm}[df]{Theorem}
\newtheorem{cor}[df]{Corollary}

\newcommand{\qed}{$\Box$}

\makeatletter \@addtoreset{equation}{section}

\newcommand{\bes}{\begin{displaymath}}
\newcommand{\ees}{\end{displaymath}}
\newcommand{\be}{\begin{equation}}
\newcommand{\ee}{\end{equation}}
\newcommand{\ba}{\begin{eqnarray}}
\newcommand{\ea}{\end{eqnarray}}
\newcommand{\bas}{\begin{eqnarray*}}
\newcommand{\eas}{\end{eqnarray*}}
\newcommand{\@Bbb}[1]{\ensuremath{\mathbb #1}}

\newcommand{\B}{{\@Bbb B}}
\newcommand{\C}{{\@Bbb C}}

\newcommand{\F}{{\@Bbb F}}
\renewcommand{\P}{{\mathbb P}}
\newcommand{\bbP}{{\P}}
\newcommand{\bbE}{{\mathbb E}}
\newcommand{\Q}{{\@Bbb Q}}
\newcommand{\bQ}{{\@Bbb Q}}
\newcommand{\N}{{\@Bbb N}}

\newcommand{\bbR}{{\@Bbb R}}
\newcommand{\W}{{\@Bbb W}}

\newcommand{\bbZ}{{\@Bbb Z}}
\newcommand{\bbT}{{\@Bbb T}}
\newcommand{\bbH}{{\@Bbb H}}

\newcommand{\la}{\lambda}

\newcommand{\ka}{\kappa}

\newcommand{\si}{\sigma}

\newcommand{\Om}{\Omega}
\newcommand{\om}{\omega}

\newcommand{\bk}{{\bf k}}

\newcommand{\bi}{{\bf i}}

\newcommand{\ep}{\varepsilon}
\newcommand{\eps}{\epsilon}

\newcommand{\@s}[1]{\ensuremath{\mathcal #1}}
\newcommand{\cA}{\@s A}
\newcommand{\cB}{\@s B}
\newcommand{\cC}{\@s C}
\newcommand{\cD}{\@s D}
\newcommand{\cE}{\@s E}
\newcommand{\cF}{\@s F}
\newcommand{\cG}{\@s G}
\newcommand{\cH}{\@s H}
\newcommand{\cI}{\@s I}
\newcommand{\cJ}{\@s J}
\newcommand{\mc}{\mathcal}
\newcommand{\cK}{\@s K}
\newcommand{\cL}{\@s L}
\newcommand{\cN}{\@s N}
\newcommand{\cM}{\@s M}
\newcommand{\cO}{\@s O}
\newcommand{\cP}{\@s P}
\newcommand{\cR}{\@s R}
\newcommand{\cS}{\@s S}
\newcommand{\cT}{\@s T}
\newcommand{\cV}{\@s V}
\newcommand{\cW}{\@s W}
\newcommand{\cX}{\@s X}
\newcommand{\cY}{\@s Y}
\newcommand{\cZ}{\@s Z}

\newcommand{\@bm}[1]{\ensuremath{\mathbf #1}}
\newcommand{\bma}{\@bm a}
\newcommand{\bmb}{\@bm b}
\newcommand{\bmc}{\@bm c}
\newcommand{\bmd}{\@bm d}

\newcommand{\bme}{\@bm e}
\newcommand{\bmf}{\@bm f}
\newcommand{\bmg}{\@bm g}
\newcommand{\bmh}{\@bm h}
\newcommand{\bmi}{\@bm i}
\newcommand{\bmj}{\@bm j}
\newcommand{\bmk}{\@bm k}
\newcommand{\bml}{\@bm l}
\newcommand{\bmm}{\@bm m}
\newcommand{\bmn}{\@bm n}
\newcommand{\bmo}{\@bm o}
\newcommand{\bmp}{\@bm p}
\newcommand{\bmq}{\@bm q}
\newcommand{\bmr}{\@bm r}
\newcommand{\bms}{\@bm s}
\newcommand{\bmt}{\@bm t}
\newcommand{\bmu}{\@bm u}
\newcommand{\bmw}{\@bm w}
\newcommand{\bmv}{\@bm v}
\newcommand{\bmx}{\@bm x}
\newcommand{\bx}{\@bm x}

\newcommand{\bmy}{\@bm y}
\newcommand{\bz}{\@bm z}

\newcommand{\by}{\@bm y}

\newcommand{\bmzero}{\@bm 0}

\newcommand{\ga}{\gamma}

\newcommand{\@g}[1]{\ensuremath{\mathfrak #1}}
\newcommand{\gA}{\@g A}
\newcommand{\gD}{\@g D}
\newcommand{\gJ}{\@g J}
\newcommand{\gF}{\@g F}
\newcommand{\gM}{\@g M}
\newcommand{\gR}{\@g R}

\newcommand{\commentout}[1]{{}}

\makeatother

\begin{document}

\title{Asymptotics of the solutions of the  stochastic  lattice   wave equation}

\author{Tomasz Komorowski\thanks{Institute of Mathematics,  UMCS,
pl. Marii Curie-Sk\l odowskiej 1,
20-031, Lublin and
IMPAN,
ul. \'{S}niadeckich 8,   00-956 Warsaw, Poland, e-mail:
komorow@hektor.umcs.lublin.pl}
\and
Stefano Olla\thanks{Ceremade, UMR-CNRS 7534, Universit\'e Paris
  Dauphine, Place Marechal Lattre de Tassigny, 75775, Paris Cedex 16, France, e-mail: olla@ceremade.dauphine.fr}
\and Lenya Ryzhik\thanks{Department of Mathematics, Stanford University,
Stanford, CA 94305, USA,
e-mail: ryzhik@math.stanford.edu}}

\maketitle

\begin{abstract}
We consider the long time limit for the solutions of a discrete wave equation with a weak stochastic forcing.
The multiplicative noise conserves the energy, and in the unpinned case also conserves the momentum. We obtain a
time-inhomogeneous Ornstein-Uhlenbeck equation for the limit wave function that holds both for square integrable and
statistically homogeneous initial data. The limit is understood in the point-wise sense in the former case, and in the weak sense
in the latter. On the other hand, the weak limit for square integrable initial data is deterministic.
\end{abstract}

\section{Introduction}

\label{intro}

Energy transport and dispersion in dynamics of oscillators in a
lattice have been investigated in many situations in order to
understand macroscopic thermal conductivity properties.
A typical example is the Fermi-Pasta-Ulam chain under the Hamiltonian
evolution corresponding to a quartic interaction potential. In the one dimension the Hamiltonian of  the closed system of
 length $N$ with periodic
boundary conditions
 is given by 
\begin{equation}
  \label{eq:fpu}
  \mc H = \sum_{y\in\bbZ/N\bbZ} \left(\frac{{\frak p}_y^2}{2m} + \frac 12
    \om_0^2 {\mathfrak q}_y^2 \right)
  + \sum_{y\in\bbZ/N\bbZ} \left[\frac {1}{2} ({\frak q}_y - {\frak
      q}_{y-1})^2 +
  {\gamma} ({\frak q}_y - {\frak q}_{y-1})^4 \right]
\end{equation}
Here $\bbZ/N\bbZ$ denotes the group $\{0,\ldots,N-1\}$ with the addition modulo $N$, ${\frak q}_y$ is the displacement of the $y$-th particle from its
equilibrium position, ${\frak p}_y$ is its momentum and $m$ is the mass. When
$\omega_0\neq 0$, the particle is confined, this breaks translation
invariance, and correspondingly the conservation of the total momentum,
 and we say that the chain is pinned. 

When $\gamma = 0$ the Hamiltonian dynamics is given by the discrete in
space linear wave equation, and the energy evolution is purely
ballistic and dispersive. If $\gamma >0$ and $\omega_0\neq 0$,   due
to the presence of the non-linearity, wave \emph{scattering} is
expected that in turn   
gives a finite thermal conductivity and consequently a diffusive
macroscopic evolution of the energy. If the chain is unpinned,
$\omega_0 =0$, and $\gamma>0$, long waves scatter rarely, giving
rise to a superdiffusive behavior of the energy \cite{sll}.   


The mathematical analysis of the macroscopic behavior of the energy is
difficult in the case of deterministic nonlinear dynamics, and recently
various models considering   stochastic perturbations of the dynamics have been proposed.
Such perturbations    generate 
scattering qualitatively similar to the one due to the
nonlinearity. 

 In order to mimic the nonlinear dynamics, a noisy perturbation we wish to consider should
  conserve energy and be local in space \cite{BO}. In the
unpinned case it is also important that it conserve the momentum, see
\cite{bborev, bbo2}. The  perturbations considered in these
papers  are given by a random exchange of momentum so that
the total kinetic energy is constant (consequently, the total
energy is preserved as well, since the position components are untouched by the noise) and the total
momentum is also conserved. This is achieved by adding, to each triple
of adjacent particles, a diffusion on the 
corresponding surface of constant energy and  momentum. Another example
of a noisy  perturbation having similar properties appears in   a
discontinuous in time model in which momenta of pairs of adjacent
particles are exchanged at independent random times that are
exponentially distributed.

When the interaction is linear, the thermal diffusivity of the energy
in these models can be explicitly computed -- it 
 is finite for the pinned model but  diverges with the size of the
 system in the unpinned case (corresponding to superdiffusive energy
 transport for the unpinned model).   

The limit dynamics for the spectral
measure of the energy in these stochastic models is investigated in \cite{BOS}, where the noise is
also rescaled in such a way that there are only finitely many
{wave collisions} in the unit macroscopic time. In a sense,
this weak noise limit is similar to the regime where phonon-Boltzmann
equation is valid in weakly nonlinear models (cf. \cite{HS}).
The dynamics is defined in the following way.
Consider the  infinite lattice $\bbZ$
with the Hamiltonian associated to the linear evolution \eqref{eq:fpu}  
($\gamma = 0$), with
$N=\infty$, perturbed by a conservative noise.       
Formally, it is given by the solution of the
stochastic differential equations:
\begin{equation}
  \label{eq:sde1}
  \begin{split}
    \dot{\frak q}_{y}(t)\ =&\ {\frak p}_y(t)\\
    d {\frak p}_y (t) \ =&\ \left(\Delta {\frak q}_y - \omega_0^2 {\frak
      q}_y\right) dt \ + d\eta_y(\eps t), 
  \end{split}
\end{equation}
where $\Delta  {\frak q}_y= {\frak q}_{y+1}+ {\frak q}_{y-1}-2 {\frak
  q}_y$ is the lattice Laplacian. The noise $d\eta_y(\eps t)$
will be added  to model random
exchange of momenta  
between the adjacent sites  so that the total  kinetic energy and momentum of the system are conserved  
(see (\ref{eq:bas}) for the precise form of the noise).
The small parameter $\eps>0$  
slows down its effect. The total Hamiltonian can be formally written as
\begin{equation}
  \label{eq:2}
  \mc H(\frak q,\frak p) =  \sum_{y\in\bbZ}\frac{{\frak p}_y^2}2 +
  \sum_{x,y \in \bbZ} \alpha_{x-y} {\frak q}_x {\frak q}_y,
\end{equation}
with $\alpha_0 = \frac 12 \omega_0^2 + 1$, $\alpha_{-1} = \alpha_{1} =- 1/2$, and $\alpha_y = 0$ otherwise. 
The dispersion relation $\om(k)$  for this system is
\begin{equation}
  \label{eq:dr0}
  \omega(k) := \sqrt{\hat\alpha(k)}=\left[\frac{\om_0^2}{2} + 2 \sin^2(\pi k)\right]^{1/2},\quad  k\in \bbT . 
\end{equation}
In fact we would admit a broader class of dispersion relations, requiring that  $\hat\alpha(k)$  is defined as in \eqref{fourier} below.
Let us introduce the complex wave function
\begin{equation}
  \label{eq:wf0}
  \psi_y(t) := (\check\omega * {\frak q})_y(t) + i {\frak p}_y(t),
\end{equation}
where $\check\omega_y$ is the inverse Fourier transform of
$\omega(k)$. Its Fourier transform  
\begin{equation}
  \label{eq:fwf0}
  \hat \psi(t,k) := \omega(k) \hat{{\frak q}}(k,t) + i \hat{\frak p}(t,k)
\end{equation}
satisfies the equation
\begin{equation}
  \label{eq:fwf01}
 d\hat \psi(t,k) = -i \omega(k) \hat \psi(t,k) dt + i  d\hat\eta(\eps t,k),
\end{equation}
where $d\hat\eta(t,k)$ 
is the Fourier transform of the noise.
 Due to the conservation properties of  the
dynamics, if the initial configuration has   finite total energy $\mc H(\frak
q(0), \frak p(0)) < +\infty$, then all the functions introduced in \eqref{eq:2} and \eqref{eq:wf0}-\eqref{eq:fwf0} are well defined and
\begin{equation*}
 {\cal H}(\frak
q(t), \frak p(t)) = \sum_y |\psi_y(t)|^2 = \int_{\bbT} |\hat\psi(t,k)|^2 dk
\end{equation*}
Therefore we can identify $|\hat\psi(t,k)|^2$ with the energy density in the mode
space. In the zero noise case,
$|\hat\psi(t,k)|^2$ is conserved for any $k\in \bbT$
(i.e. $\partial_t |\hat\psi(t,k)|^2 =0$).
The stochastic conservative perturbation mixes the energies between 
different modes $k$, and $|\hat\psi(t,k)|^2$ becomes a random variable. 
The evolution of the average energy $\mc E(t,k) :=
\mathbb E|\hat\psi(t,k)|^2$ was considered in~\cite{BOS}. Since the stochastic perturbation is
of order $\epsilon$, to have a visible effect of mixing of
different modes we have to look at the time scale $\epsilon^{-1}t$.
It was shown in \cite{BOS}  that the limit
\begin{equation}
  \label{eq:bos0}
  \lim_{\eps\to 0} \mc E\left(\frac t\eps , k\right) = \bar{\mc
  E}\left( t , k\right)  
\end{equation}
exists in the sense of distributions, and 
is the solution of the linear kinetic equation
\begin{equation}
  \label{eq:phbolhom}
  \partial_t \bar{\mc E}\left( t , k\right) = 
  \int_{\bbT} R(k,k') \left[\bar{\mc E}\left( t , k'\right) -\bar{\mc
    E}\left( t , k\right)  \right] dk'
\end{equation}
with the initial condition $\bar{\mc E}\left(0 , k\right) = |\hat\psi(0,k)|^2$. 
The scattering kernel $R(k,k')$ is
given by \eqref{kernel}
below.

The goal of the present article  is to obtain  a direct information on the wave function
$\hat\psi(t/\eps,k)$, as was done in~\cite{BKRschr} for the Schr\"odinger equation, and not only for the
average energy. 
It follows from \eqref{eq:fwf01} that  the unperturbed (by noise) evolution of this function
is governed by the highly oscillating factor
$e^{-i\om(k)t/\ep}$ (after we rescale the time).  It is therefore reasonable to consider, in case of the perturbed system,  the compensated wave function of the form 
$$
\tilde\psi^{(\eps)}(t,k):=e^{i\om(k) t/\ep} \hat\psi(t/\ep,k).
$$
We show that once we compensate for  fast oscillations,  
the wave function converges in law to  the solution a  Langevin equation driven by \eqref{eq:phbolhom}. 
More precisely, we prove in Theorem \ref{main-thm1} below,  existence of
the limit (in law and pointwise in $k$): 
\begin{equation}
    \label{eq:comp0}
    \lim_{\ep\to 0} \tilde\psi^{(\eps)}(t,k) = \tilde\psi(t,k).
  \end{equation}
The limit $\tilde\psi(t,k)$ is a complex valued stochastic process
satisfying the linear (time inhomogeneous) Ornstein-Uhlenbeck equation
  \begin{equation}
    \label{eq:lan0}
    d \tilde\psi(t,k) = - \frac{\hat\beta(k)}{4} \tilde\psi(t,k) dt 
    + \sqrt{\mc R(t,k)} dw_k(t),
\end{equation}
with the initial condition $\tilde\psi(0,k) = \hat\psi(0,k)$.
Here 
  \begin{equation}
  \label{R-B}
  \hat\beta(k) =2\int_{\bbT}R(k,k')dk'
  \end{equation}
\begin{equation}
  \label{eq:beta0}
  \mc R(t,k) = \int_{\bbT} \bar{\mc E}(t, k') R(k,k') dk',
\end{equation}
and $\{w_k(t)\}$ is a family of pairwise independent standard complex valued Brownian motions 
parametrized by $k\in\bbT$. That is, they are   complex valued, jointly Gaussian, centered
processes satisfying
$$\mathbb E[w_k(t) w_{k'}(s)] = 0\quad\mbox{and}\quad 
\mathbb
E[w_{k'}^*(t) w_k(s)] =\delta_{k,k'} t\wedge s
$$ for all $t,s\ge0$ and $k,k'\in\bbT$. Here
$\delta_{k,k'}=0$ for $k\not=k'$ and $\delta_{k,k}=1$.
Equation \eqref{eq:lan0} has the explicit solution
\begin{equation}
  \label{eq:3}
  \tilde\psi(t,k) = e^{-\frac 14\hat\beta(k)t} \hat\psi(0,k) + \int_0^t
  e^{-\frac 14\hat\beta(k)(t-s)} \sqrt{\mc R(s,k)} dw_k(s).
\end{equation}
In particular, we have
\begin{equation*}
  \mathbb E |\tilde\psi(t,k)|^2= 
  e^{-\frac12\hat\beta(k)t} |\hat\psi(0,k)|^2 + \int_0^t
  e^{-\frac12\hat\beta(k)(t-s)} \mc R(s,k) ds
\end{equation*}
which is equivalent to \eqref{eq:phbolhom}, since $\bar{\mc
  E}(t,k) = \mathbb E |\tilde\psi(t,k)|^2$.
Initial conditions such that $\int_{\bbT} |\hat\psi(0,k)|^2 dk < \infty$
correspond to a \emph{local} perturbation of the zero temperature
equilibrium. We are also interested in the macroscopic evolution
of the equilibrium states at a positive temperature $T >0$, 
starting with  random data distributed by the Gibbs
measure at temperature $T$. In the mode space this is a centered,
complex valued,  Gaussian random field  with distribution valued $\hat\psi(k)$. Its covariance is given by
\begin{equation}
  \label{eq:gibbscov}
  \bbE[\hat\psi^*(k) \hat\psi(k')] =T \delta(k-k'), \qquad 
   \bbE[\hat\psi(k) \hat\psi(k')] = 0.
\end{equation}
Here $\delta(k-k')$ is Dirac's delta function.
For any $T$, the corresponding Gibbs measure is invariant under the
dynamics, due to the conservation of energy.  
Actually, in Section \ref{sec2.3.2} we consider more general 
class of space
homogeneous Gaussian random initial conditions whose law is not
necessarily stationary in time. More precisely, we show (see Theorem
\ref{main-thm2}) that if the law of the initial 
condition is a homogeneous, centered Gaussian field with the covariance given by
\begin{equation*}
   \bbE\left[\hat\psi(k)^* \hat\psi(k')\right] = \mc E_0(k)\delta(k-k')  , \qquad 
   \bbE\left[\hat\psi(k) \hat\psi(k')\right]= 0,
\end{equation*}
then  the compensated wave function   
converges in law,   as a continuous in time process taking values in an appropriate distribution space, to the solution of the time
inhomogeneous stochastic equation:
\begin{equation}
  \label{eq:spdet1}
    d \tilde\psi(t,k) = - \frac{\hat\beta(k)}{4} \tilde\psi(t,k) dt 
    + \sqrt{\mc R(t,k)} dW(t,k).
\end{equation}
Here, $\mc R(t,k) $ is given by \eqref{eq:beta0} and $\bar{\mc
E}(t,k)$ is the solution of the deterministic equation \eqref{eq:phbolhom}
with the initial condition $\bar{\mc E}(0,k) = \mc E_0(k)$, while $d W(t,k)$ is a white noise on $\bbR
\times \bbT$, a complex valued Gaussian process with the covariance  
 $$
\mathbb E[dW(t,k)d W^*(s,k')] = \delta(k-k') \otimes\delta(t-s)dtds
$$ 
and ${\cal R}(t,k)$ is given by \eqref{eq:beta0}.
The solution of  \eqref{eq:spdet1} is also explicit: 
$\tilde\psi(t)$ is the distribution 
\begin{equation*}
\tilde\psi(t) = e^{-\hat \beta t/4}\hat\psi  +  \int_0^t  e^{-\hat \beta(t-s)/4}\mc R^{1/2}(s) dW(s) .
\end{equation*}
In particular, in  the case of the initial condition distributed according to a Gibbs measure,
 the solution $\hat\psi(t,k)$ of \eqref{eq:fwf01} has the same law for all times, therefore 
$\bar{\mc E}(t,k) = T$ for all $t\ge0$.  In this case, \eqref{R-B} shows that  ${\cal R}(t,k)=\hat\beta(k)T/2$.
Therefore, as a consequence of \eqref{eq:spdet1},    the limit of  the compensated wave function is the solution of the linear
infinite dimensional stochastic differential equation:
\begin{equation}
  \label{eq:spde0}
    d \tilde\psi(t,k) = - \frac{\hat\beta(k)}{4} \tilde\psi(t,k) dt 
    +\sqrt{\frac{T \hat\beta(k)}{2}} dW(t,k).
\end{equation}
In the general case, when $\mc E_0(k)$ is not constant, we have
$$\lim_{t\to\infty} \bar{\mc E}(t,k) = \int_{\bbT} \mc
E_0(k') dk' = T,
$$ 
hence, equation \eqref{eq:spde0} describes the asymptotic
stationary regime of \eqref{eq:spdet1} where the temperature is given
by the average of the initial energy over all the modes $k$. 
Recall that the microscopic noise conserves the
total energy and that the resulting temperature $T$ depends only on
the law of the initial condition.

Let us also comment on the difference between the square integrable   and distribution-valued initial data. 
While the  
Ornstein-Uhlenbeck  equations (\ref{eq:lan0}) and (\ref{eq:spdet1}) look similar, there are some important differences between them.
The noises appearing in (\ref{eq:lan0}) are all of size $1$ and mutually independent  for different $k$-s, while the noise appearing in (\ref{eq:spdet1}) is $\delta$-correlated in $k$. As a result the solution
of the first equation is an ensemble of mutually independent time inhomogeneous one dimensional Ornstein-Uhlenbeck processes. On the other hand, in the case of  (\ref{eq:spdet1}) the resulting distribution valued Ornstein-Uhlenbeck process is $\delta$-correlated in $k$. In addition,
for the square integrable data, the limit equation holds point-wise in $k$. If one considers the limit in the sense of 
distributions (that is, integrated against a test function) for such initial data, the stochasticity
is removed, due to the fact that independent random variables, representing the solution for different modes, are simply averaged out (via the law of large numbers).
As a result the limit is described simply
by attenuation of the initial condition by an exponential factor $e^{-\beta(k)t/4}$ (see part (ii) of Theorem \ref{main-thm1}) -- that is,
by (\ref{eq:lan0}) with no stochastic forcing. 
This result stands in sharp contrast with the case of spatially homogeneous initial data (note that then the energy has to be infinite) when the respective limit in the sense of distributions is stochastic, see \eqref{eq:spdet1}, and fluctuations can not be averaged out by integration in $k$.

Finally, we note that the sole reason why we restrict ourselves to the case of one dimensional integer lattice is to avoid excessive complication of the notation that could obscure the main points of the argument. The technique of  our proof can be straightforwardly applied in the case of lattice $\bbZ^d$. The dynamics of the corresponding perturbed system is given then by equation (45) of \cite{BOS} and our results contained in Theorems \ref{main-thm1}   and Theorem \ref{main-thm2} can be easily adjusted to deal with the case of a multidimensional lattice.

The paper is organized as follows.  Section~\ref{prelim} contains the precise mathematical formulation of the problem and necessary 
definitions.
We formulate the results for the convergence of compensated wave function in 
 Section~\ref{sec3}, see Theorem \ref{main-thm1}  for square integrable initial data, and Theorem \ref{main-thm2} for
spatially homogeneous, Gaussian initial distributions.
The proofs of these results are presented in Sections \ref{sec4} and \ref{sec5}, respectively.

{\bf Acknowledgement.}
T.K. acknowledges the support of Polish Ministry of Higher Education
grant NN201419139,  S.O. acknowledges the support by the ERC AdG
246953 (MALADY) and by ANR-10-BLAN 0108 (SHEPI), L.R. acknowledges the support
by NSF grant DMS-0908507. This work was also supported by NSSEFF fellowship by AFOSR.

\section{Preliminaries}

\label{prelim}

\subsection{Infinite system of interacting harmonic oscillators}

The dynamics of the system of oscillators  can be written formally  as
a system of  It\^o stochastic differential equations indexed by $y\in\mathbb Z$
\begin{eqnarray}
d{\frak q}_{y}(t) &=&{\frak p}_y(t)dt
\label{eq:bas}\\
&&\nonumber\\
 d{\frak p}_y(t) &=& - (\alpha*{\frak q}(t))_y\ dt-\frac{\eps}{2}(\beta*{\frak p}(t))_y\ dt
+\sqrt{\eps}\sum_{z=-1,0,1}(Y_{y+z}{\frak p}_y(t))dw_{y+z}(t).\nonumber 
\end{eqnarray}
Here 
$$
Y_x:=({\frak p}_x-{\frak p}_{x+1})\partial_{{\frak p}_{x-1}}+({\frak p}_{x+1}-{\frak p}_{x-1})\partial_{{\frak p}_{x}}+({\frak p}_{x-1}-{\frak p}_{x})\partial_{{\frak p}_{x+1}}
$$
and
  $\{w_y(t),\,t\ge0\}$, $y\in\bbZ$ is a family of  i.i.d. one dimensional, real valued, standard Brownian motions,  
  that are  non-anticipative over the filtered probability space $(\Om,{\cal F},\{{\cal F}_t\},\bbP)$. In addition,
 \begin{equation}
 \label{040910}
\beta_y=\Delta\beta^{(0)}_y:=\beta^{(0)}_{y+1}+\beta^{(0)}_{y-1}-2\beta^{(0)}_y
\end{equation}
with 
$$
 \beta^{(0)}_y=\left\{
 \begin{array}{rl}
 -4,&y=0\\
 -1,&y=\pm 1\\
 0, &\mbox{ if otherwise.}
 \end{array}
 \right.
 $$
   Recall that the lattice Laplacian of  $g:\bbZ\to\mathbb C$ is given by
  $\Delta g_y:=g_{y+1}+g_{y-1}-2g_y$.  
  
  To understand why we choose this particular stochastic perturbation of the
  Hamiltonian dynamics, let us observe that we want a (continuous)
  noise acting only on the velocities, as local as possible, but
  conserving total momentum and kinetic energy. 
   This explains why, given a site $y$, only the momenta at sites
   $y+z$, $z=-1,0,1$  are exchanged randomly. For that reason we
   consider the vectors $Y_{x}$ that are tangent to   
 the local energy and momentum surfaces  
\begin{equation}
\label{051207a}
{\frak p}_{x-1}^2+{\frak p}_{x}^2+{\frak p}_{x+1}^2\equiv {\rm const}
\end{equation} 
and 
\begin{equation}
\label{051207b}
{\frak p}_{x-1}+{\frak p}_{x}+{\frak p}_{x+1}\equiv {\rm const}.
\end{equation} 
The  SDE \eqref{eq:bas} defines a Markov process whose (formal) generator is given by
\begin{equation}
  \label{eq:gen}
  L = A + \eps S, \qquad S=\frac12 \sum_x Y_x^2,
\end{equation}
where $A$ is the Hamiltonian vector field given by the usual Poisson
brackets with the Hamiltonian.
In particular $-(\beta*p)_y/2 = S p_y$.

The Fourier transform of  a  square integrable sequence of complex numbers $\{\gamma_y,\,y\in\bbZ\}$ is defined  as
  \begin{equation}
  \label{fourier}
  \hat \gamma(k)=\sum_{y\in\bbZ}\gamma_ye_y(k), \quad k\in\bbT.
\end{equation}
Here
$$
e_y(k):=\exp\{-i2\pi yk\}, \quad y\in\bbZ
$$ is  the standard orthonormal base in $L^2(\bbT)$.
The one dimensional torus $\bbT$  considered in this article  is understood as the interval $[-1/2,1/2]$ with identified endpoints.
The  inverse transform is given by
\begin{equation}
\label{inv-fourier}
\check f_y=\int_{\bbT} f(k)e^*_y(k)dk
, \quad y\in \bbZ
\end{equation} for any $f$ belonging to $L^2(\bbT)$ - the space of complex valued, square integrable functions.
  A simple calculation shows that
\begin{equation}
\label{beta}
\hat \beta(k)=8\sin^2(\pi k)\left[1+2\cos^2(\pi k)\right].
\end{equation}

We assume also (cf  \cite{BOS}) that
 \begin{itemize}
 \item[a1)] $\{\alpha_y,\,y\in\bbZ\}$ is real valued and there exists $C>0$ such that $|\alpha_y|\le Ce^{-|y|/C}$ for all $y\in \bbZ$,
  \item[a2)] $\hat\alpha(k)$ is also real valued 
and  $\hat\alpha(k)>0$ for $k\not=0$ and in case $\hat \alpha(0)=0$ we  have  $\hat\alpha''(0)>0$.
 \end{itemize}
   The above conditions imply that both functions $y\mapsto\alpha_y$ and $k\mapsto\hat\alpha(k)$ are even. In addition, $\hat\alpha\in C^{\infty}(\bbT)$ and in case $\hat\alpha(0)=0$ we have  $\hat\alpha(k)=\sin^2(\pi k)\phi(k)$ for some strictly positive even function  $\phi\in C^{\infty}(\bbT)$.
Recall that the function $\om(k):=\sqrt{\hat \alpha (k)}$ is { the
  dispersion relation}.

\subsection{Evolution of the wave function}
%
 
For a given  $m\in\bbR$ we define the space $H^m(\bbT)$ as the completion of
$C^\infty(\bbT)$ under the norm
$$
\|f\|^2_{H^m(\bbT)}:=\sum_{y\in\bbZ}(1+y^2)^m|\check  f_y|^2.
$$
We shall denote by $\langle\cdot,\cdot\rangle$ the scalar product on $L^2(\bbT)$. By continuity it extends in an obvious way  to $H^m(\bbT)\times H^{-m}(\bbT)$ for an arbitrary $m\in\bbR$.

It is convenient to introduce the wave function that, adjusted to the macroscopic time, is given by
\begin{equation}
\label{011307}
\psi^{(\eps)}(t):=\check {\om} * {\frak q}\left(\frac{t}{\eps}\right)+i{\frak p}\left(\frac{t}{\eps}\right).
\end{equation}
Here $\{\check  \om_y,\,y\in\bbZ\}$ is the inverse Fourier transform of   
\begin{equation}
\label{om}
\om(k):=\sqrt{\hat \alpha (k)}.
\end{equation}
 We shall consider the Fourier transform of the wave function
\begin{equation}
\label{011307a}
\hat\psi^{(\eps)}(t,k):=\om(k)\hat {\frak q}\left(\frac{t}{\eps},k\right)+i\hat{\frak p}\left(\frac{t}{\eps},k\right).
\end{equation}
Using \eqref{eq:bas} as a motivation, we obtain formally,  by considering the Fourier transform of  \eqref{eq:bas}, that 
  \begin{eqnarray}
 \label{basic:sde:2}
&&
 d\hat\psi^{(\eps)}(t)=A[\hat\psi^{(\eps)}(t)]dt +Q[\hat\psi^{(\eps)}(t)]dW(t),\\
 &&
\hat\psi^{(\eps)}(0)= \hat\psi,\nonumber
 \end{eqnarray}
 where   $ \hat\psi\in L^2(\bbT)$, and mapping $A:L^2(\bbT)\to L^2(\bbT)$ is defined by 
 \begin{equation}
\label{040607}
 A[f](k):=-\frac{i}{\eps} \om(k)f(k)- \frac{\hat\beta(k)}{4} [f_{1}(k)-f_{-1}(k)],\quad\forall\,f\in  L^2(\bbT).
 \end{equation}
 Here
  \begin{equation}
   \label{020906}
 f_1(k):=f(k)\quad\mbox{and}\quad f_{-1}(k):=f^*(-k).
\end{equation}
In addition, 
 $Q[g]:L^2(\bbT)\to L^2(\bbT)$ is a linear mapping that for any $g\in L^2(\bbT)$ is given by
\begin{equation}
\label{053009}
 Q[g](f)(k):=i\int_{\bbT}r(k,k')[g_{1}(k-k')-g_{-1}(k-k')]f(k')dk',\quad\forall\,f\in  L^2(\bbT),
\end{equation}
where
 \begin{eqnarray*}
 &&r(k,k'):=\sin(2\pi k)+\sin[2\pi(k-k')]+\sin[2\pi(k'-2k)]
 \\
 &&~~~~~~~~~~=4\sin(\pi k)\sin[\pi (k-k')]\sin\left[(2k-k')\pi\right]
 ,\quad k,k'\in \bbT.
 \end{eqnarray*}
The cylindrical Wiener process on $L^2(\bbT)$ appearing in \eqref{basic:sde:2} is  $dW(t):=\sum_{y\in\bbZ} e_ydw_y(t)$.

It can be easily checked that $\sum_{y\in\bbZ}\|Q[g](e_y)\|_{L^2(\bbT)}^2\le C\|g\|_{L^2(\bbT)}^2$ for some $C>0$ and  all $g\in L^2(\bbT)$ so $Q[g]$ is Hilbert-Schmidt, which ensures that
 $$
Q[\hat\psi^{(\eps)}(t)]dW(t):=\sum_{y\in\bbZ} Q[\hat\psi^{(\eps)}(t)](e_y)dw_y(t)
$$
is summable in $L^2(\bbT)$, both in the $L^2$ and a.s. sense. It is also obvious that the mapping $A$ is Lipschitz.
 Using   Theorem 7.4, p. 186, of \cite{DZ}  one concludes therefore that there exists  an $L^2(\bbT)$-valued, adapted process $\{\hat\psi^{(\eps)}(t),\,t\ge0\}$ that is a unique solution to \eqref{basic:sde:2}. 
In addition, see Section 2 of \cite{BOS},  the total energy is conserved:
\begin{equation}
\label{conservation}
\|\hat\psi^{(\eps)}(t)\|_{L^2(\bbT)}={\rm const},\quad\forall\,t\ge0
\end{equation}
 for a.s. realization of Brownian motions  and an initial condition  from $L^2(\bbT)$.

 \subsection{Compensated wave function}
 
 \label{pseudo-wigner}

 Let us define the compensated wave function
 $$
 \tilde \psi^{(\eps)}(t,k):=\hat \psi^{(\eps)}(t,k)\exp\left\{it\frac{\om(k)}{\eps}\right\}.
 $$
 From \eqref{basic:sde:2} we obtain
the following equation 
\begin{eqnarray}
\label{mollified-eqt}
&&d \tilde \psi^{(\eps)}(t,k)=
 {\cal A}\left[\frac{t}{\eps},\tilde\psi^{(\eps)}(t)\right](k)dt +d\tilde{\cal M}^{(\eps)}_t(k),\nonumber\\
 &&
\tilde\psi^{(\eps)}(0)= \hat\psi,
 \end{eqnarray}
where $ \hat\psi\in L^2(\bbT)$, $ {\cal A}[t,\cdot]:L^2(\bbT) \to L^2(\bbT)$ 
 \begin{equation}
\label{012808}
 {\cal A}[t,f](k):=-\frac{\hat\beta(k)}{4}\left[ f(k)-\exp\left\{2i\om(k)t\right\}f^*(-k)\right].
 \end{equation}
 The martingale term equals
\begin{equation}
\label{060410}
d \tilde{\cal M}^{(\eps)}_t:=\tilde Q\left[\frac{t}{\eps},\tilde\psi^{(\eps)}(t)\right]dW(t),
\end{equation}
where for any $g\in  L^2(\bbT)$ and $t\ge0$, the operator  $\tilde Q[t,g] :L^2(\bbT)\to L^2(\bbT)$, is given by
 \begin{equation}
\label{022808}
\tilde Q[t,g](f)(k):=i\sum_{\si=\pm1}\si\int_{\bbT}r(k,k')g_{\si}(k-k')f(k')  
\exp\left\{i[\om(k)-\si\om(k-k')]t\right\}dk'.
  \end{equation}
Using a standard  theory of S.P.D.E.-s, see \cite{DZ},  we can show
the following result. 
\begin{prop}
\label{prop010910}
Suppose that $-3/2<m<1$.
If the initial condition $\hat \psi(\cdot)$ belongs to $H^{m}(\bbT)$
then  there exists a unique solution $(\tilde \psi^{(\eps)}(t))$ of
\eqref{mollified-eqt} in  $H^{m}(\bbT)$.   
\end{prop}
The proof of this result shall be presented in Appendix
\ref{appA}. Since the dispersion relation $\om(\cdot)$ might not be differentiable in the classical sense at $0$ (but it belongs to $H^1(\bbT)$) we cannot guarantee better  regularity of the solutions of \eqref{mollified-eqt}. Recall that the classical Sobolev embedding theorem
ensures that $H^m(\bbT)$, for $m>1/2$, is embedded in the space of
continuous functions on the torus $C(\bbT)$, see e.g. Theorem 7.10,
p. 155 of \cite{GT}.

\section{Convergence of the compensated process}\label{sec3}
\subsection{Square integrable initial data }

\label{sec3.1}


Before formulating the result we introduce some auxiliaries. First, 
for any $k_1,k_2\in \bbT$ let us denote
  $$
{\cal K}(k_1,k_2)=\bigcup_{\si_1,\si_2,\si_3=\pm1}
 [k:\om(k_1)+\si_3\om(k-k_1)=\si_1[\om(k_2)+\si_2\om(k-k_2)]]
 $$
We shall require that:
\[
\hbox{ {\bf Condition} $\om$)
for any $k_1\not=k_2$ the one dimensional Lebesgue measure  $m_1({\cal
  K}(k_1,k_2))=0$.  
}
\]
More detailed discussion of this condition shall be carried out in Remark 2 after Theorem \ref{main-thm1} below.

Define the scattering operator ${\cal L}:L^1(\bbT)\to L^1(\bbT)$ by  
\begin{equation}
\label{L}
{\cal L}f(k):=\int_{\bbT}R(k,k')[f(k')-f(k)]dk',\quad f\in L^1(\bbT),
\end{equation}
where the scattering kernel is given by
 \begin{eqnarray}
 \label{kernel}
 &&
\!\!\!\!\!\!\!\!R(k,k'):=r^2(k,k-k')+r^2(k,k+k')\\
&&
=16\sin^2(\pi k)\sin^2(\pi k')\left\{\sin^2\left[\pi (k+k')\right]+\sin^2\left[\pi (k-k')\right]\right\}.\nonumber
 \end{eqnarray}
Suppose that $ \hat\psi\in L^2(\bbT)$. Let
\begin{equation}
\label{010810}
{\cal R}(t,k):=\int_{\bbT}R(k,k')\bar{\cal E}(t,k')dk',
\end{equation}
where $\bar{\cal E}(t,k)$ is the unique  solution in $C(\bbR,L^1(\bbT))$ of an equation
\begin{equation}
\label{100510}
\bar{\cal E}(t,k)= |\hat\psi(k)|^2+\int_0^t{\cal L}\bar{\cal E}(s,k)ds.
\end{equation}
The existence and uniqueness of solutions in \eqref{100510} follows from the fact that ${\cal L}$ is clearly a bounded operator on $L^1(\bbT)$.
The solution then is given by $\bar{\cal E}(t)=P^t\bar{\cal E}(0)$, where $\bar{\cal E}(0):=|\hat\psi|^2$ and $(P^t)$ is the contraction semigroup  on $L^1(\bbT)$ generated by ${\cal L}$. 

Assume also that  $\{w_k(t),\,t\ge0\}$ is a family of pairwise independent standard, one dimensional,
complex valued Brownian motions indexed by $k\in\bbT$. 
Our first principal result can be stated as follows.
\begin{thm}
\label{main-thm1}

Suppose that the dispersion relation $ \om(\cdot)$ satisfies condition $\om)$. Then, the following are true:

(i) if $ \hat\psi\in H^{m}(\bbT)$ for some $m>1/2$ then there exists a solution $\tilde \psi^{(\eps)}(t)$ of \eqref{mollified-eqt} that belongs a.s. to $C(\bbT)$ for all $t\ge 0$. In addition, given an integer $n\ge1$ and  $k_1,\ldots,k_n\in\bbT$,   the processes  \linebreak
$\{
(\tilde\psi^{(\eps)}(t,k_1),\ldots,\tilde\psi^{(\eps)}(t,k_n)),\,t\ge0\}$   converge in law over $C([0,+\infty);\mathbb C^n)$, as
$\eps\to0+$,   to  $\{(\tilde\psi(t,k_1),\ldots,\tilde\psi(t,k_n)),\,t\ge0\}$, where $\{\tilde\psi(t,k),\,t\ge0\}$ is  a complex valued, non-homogeneous in time
Ornstein-Uhlenbeck process that is the solution of the equation 
\begin{eqnarray}
\label{limit-eqt1}
d \tilde\psi(t,k)&=&
 -\frac{\hat\beta(k)}{4}\tilde \psi(t,k)dt +{\cal R}^{1/2}(t,k)dw_k(t),\nonumber\\
 \tilde \psi(0,k)&=& \hat\psi(k),
 \end{eqnarray}

(ii)
if  $ \hat\psi\in L^2(\bbT)$, then for any $f\in  L^2(\bbT)$ and $t_*>0$  we have
\begin{equation}
\label{040510}
\lim_{\eps\to0+}\sup_{t\in[0,t_*]}\left|\langle\tilde \psi^{(\eps)}(t)-\bar \psi(t),f\rangle\right|=0
\end{equation}
in probability. Here $\bar \psi(t)$ is given by 
 \begin{equation}
 \label{012909}
  \bar \psi(t,k):=\hat \psi_0(k)\exp\left\{-\frac{t\hat\beta(k)}{4}\right\}.
 \end{equation}
\end{thm}
{\bf Remark 1.} 
We claim that 
 \begin{equation}
\label{020910}
\lim_{t\to+\infty}\sup_{k\in\bbT}|{\cal R}(t,k)- (\hat\beta(k)/2)T|=0,
 \end{equation}
 where $T=\|\hat\psi_0\|_{L^2(\bbT)}^2$. The above easily follows 
 from \eqref{010810}, provided we show that any solution $\bar{\cal
   E}(t,k)$ of \eqref{100510} satisfies 
 \begin{equation}
\label{010910}
\lim_{t\to+\infty}\|\bar{\cal E}(t)- T\|_{L^1(\bbT)}=0.
 \end{equation}
To prove \eqref{010910} recall that operator ${\cal L}$ given by \eqref{100510} is a generator of  a strongly continuous semigroup $(P^t)$ of contractions on $L^1(\bbT)$. 
 In fact, it  is also a semigroup of contractions when restricted to
 any $L^p(\bbT)$, for  $1\le p\le+\infty$,  
 strongly continuous, provided that $p\in[1,+\infty)$.  
When $p=2$ generator ${\cal L}$ is symmetric (and so is each $P^t$) and 
$$
\langle {\cal L} f,f\rangle=-\frac12\int_{\bbT^2}R(k,k')|f(k')-f(k)|^2dkdk'\le0,\quad  \forall \,f\in L^2(\bbT). 
$$
Hence $0$ is a simple eigenvalue of ${\cal L}$ in $L^2(\bbT)$, i.e. if
$f\in L^2(\bbT)$ and satisfies ${\cal L}f=0$, then $f$ is a constant. 
This immediately implies that for $\bar{\cal E}(0)\in L^2(\bbT)$ with
 $T:=\int_{\bbT}\bar{\cal E}(0,k)dk$ we have 
 \begin{equation}
\label{010910a}
\lim_{t\to+\infty}\|\bar{\cal E}(t)- T\|_{L^2(\bbT)}^2=\lim_{t\to+\infty}\int_0^{+\infty}e^{-\la t}\mu(d\la)=0,
 \end{equation}
 where $\mu$ is the spectral measure of $\bar{\cal E}(0)-T$ corresponding to ${\cal L}$. This in particular implies \eqref{010910} in case the initial data  is square integrable. If $\bar{\cal E}(0)$ only belongs to $L^1(\bbT)$ we obtain \eqref{010910} approximating first $\bar{\cal E}(0)$  by square integrable functions and then using \eqref{010910a} together with the fact that $(P^t)$ is a contraction semigroup on $L^1(\bbT)$.

From \eqref{020910} we obtain, for any $k\in \bbT$,
\begin{equation}
\label{040510a}
\lim_{t\to +\infty}\bbE\left|\tilde \psi(t,k)-\tilde \psi_s(t,k)\right|^2=0,
\end{equation}
where $ \tilde\psi_s(t,k)$ is a  time homogeneous Ornstein-Uhlenbeck process given by
\begin{eqnarray}
\label{limit-eqt1a}
d \tilde\psi_s(t,k)&=&
 -\frac{\hat\beta(k)}{4}\tilde \psi_s(t,k)dt +\sqrt{\frac{\hat\beta(k)T}{2}}dw_k(t),\nonumber\\
 \tilde \psi_s(0,k)&=& \hat\psi(k).
 \end{eqnarray}

{\bf Remark 2.} Let us also  comment briefly on condition $\om)$. A similar hypothesis appears in the wave turbulence theory  under the name of  a {\em no resonance condition}, see e.g. \cite{lfz}. 
%
%
%
In our context  we use it, among others, to prove the asymptotic (in the limit $\eps\to 0+$)
independence of $\tilde\psi^{(\eps)}(t,k)$ for different $k$. This independence implies, in particular,
the self-averaging property of the energy $ |\tilde \psi^{(\eps)}(t,k)|^2$  i.e. its convergence in probability
 to  a deterministic limit, as $\eps\to0+$, in the weak topology, see Proposition \ref{lm013108} below. 
This observation plays a crucial r\^ole  in the proof of part (i) of Theorem \ref{main-thm1}.
Without lack of resonance  condition of the type $\om$), it is  in principle possible that the second mixed moment of the energy corresponding to different modes 
  does not vanish in the limit, as $\eps\to0+$, so that the key estimate \eqref{phi} below fails making self-averaging of   energy impossible.
%

%

The following simple criterion is useful for verification of condition
$\om)$, e.g. for dispersion relation $\om(k)$ of the
form~\eqref{eq:dr0}. Recall that from the assumptions made we know
that $\om\in C^{\infty}(\bbT\setminus\{0\})$. 
\begin{lm}
\label{lmK}
Suppose that  the dispersion relation $\om(\cdot)$   satisfies the following condition: for any $|a|<1/2$ and $\si=\pm1$ the set of solutions of an equation
\begin{equation}
\label{sec-der}
\om'(k)=\si\om'(k+a)
\end{equation}
is possibly of positive Lebesgue measure in $\bbT$, only if $a=0$ and $\si=1$.
Then, for any $(k_1,k_2)$ such that $k_1\not=k_2$  the hypothesis $\om)$ holds.
\end{lm}
{\bf Proof.} Fix $(k_1,k_2)$ such that $k_1\not=k_2$.  To simplify we consider only the set ${\cal K}_1$ that corresponds to  $\si_1=\si_2=\si_3=1$ and  prove that:
$$
{\cal K}_1(k_1,k_2):= [k:\om(k_1)+\om(k-k_2)=\om(k_2)+\om(k-k_1)]
$$
is of null Lebesgue measure.
The remaining cases can be dealt with similarly.  
 Suppose, on the contrary, that the Lebesgue measure of the set is positive. 
Then almost every  point of ${\cal K}_1(k_1,k_2)$ is a density point
of the set. In particular that means that at any such point 
we have
$$
\om'(k-k_2)=\om'(k-k_1)
$$
but this would clearly contradict the assumption made in the statement of the  lemma.
 \qed

It is quite straightforward to verify that  the above lemma applies to the dispersion relation of the form \eqref{eq:dr0}. 

%
%

\subsection{Statistically homogeneous initial data }

\label{sec2.3.2}
For a given non-negative $m$ we assume that  the initial data $\hat \psi$ is an  $H^{-m}(\bbT)$ 
valued Gaussian random element. 
 More precisely, suppose that ${\cal E}_0(\cdot)$ is a non-negative function such that
 \begin{equation}
 \label{011210a}
 \sum_{x\in\bbZ}|\langle {\cal E}_0,e_x\rangle|<+\infty,
 \end{equation}
  $\{\xi_y,\,y\in\bbZ\}$ are i.i.d. complex Gaussian random variables such that $\bbE\xi_0=0$ and $\bbE|\xi_0|^2=1$, and
\begin{equation}
\label{053110}
\hat\psi(k)=\sum_{y\in\bbZ}\xi_y {\cal E}^{1/2}_0(k)e_y(k).
\end{equation}
The law of $\hat \psi$ is supported in $H^{-m}(\bbT)$, provided that $m>1/2$. 
 Its covariance form equals
\begin{equation}
\label{021310aa}
{\cal C}(J_1,J_2):=\bbE\left[\langle J_1,\hat\psi\rangle\langle J_2,\hat\psi\rangle^* \right]=\int_{\bbT}{\cal E}_0(k)J_1(k)J_2^*(k)dk
\end{equation}
for any $J_1,J_2\in C^\infty(\bbT)$. The Gibbs equilibrium states described in the introduction correspond to 
${\cal E}_0(k)\equiv \hbox{const}$.
%
Using Proposition \ref{prop010910} we conclude that  equation \eqref{mollified-eqt} has a unique mild solution $\{\tilde \psi^{(\eps)}(t),\,t\ge0\}$ whose realizations belong to $C([0,+\infty);H^{-m}(\bbT))$, provided $m<3/2$.

Let  ${\cal R}(t,k)$ be given by \eqref{010810} with $\bar{\cal E}(t,k)$  the solution 
of \eqref{100510} satisfying $\bar{\cal E}(0,k)={\cal E}_0(k)$.
Observe that the operator 
$
f(k)\mapsto {\cal R}^{1/2}(t,k)f(k)
$
is Hilbert-Schmidt, when considered from $L^2(\bbT)$ to  $H^{-m}(\bbT)$, provided $m>1/2$.
Indeed
$$
\sum_y\|{\cal R}^{1/2}(t)e_y\|_{H^{-m}(\mathbb T)}^2=\sum_{y,y_1}(1+y_1^2)^{-m}\left|\int{\cal R}^{1/2}(t,k)e_{y-y_1}(k)dk\right|^2.
$$
By Plancherel's identity the right hand side equals
\begin{eqnarray*}
&&
\sum_{y_1}(1+y_1^2)^{-m}\sum_{z}\left|\int{\cal R}^{1/2}(t,k)e_{z}(k)dk\right|^2=\sum_{y_1}(1+y_1^2)^{-m}\|{\cal R}^{1/2}(t,\cdot)\|^2_{L^2(\mathbb T)}\\
&&
=\sum_{y_1}(1+y_1^2)^{-m}\|{\cal R}(t,\cdot)\|^2_{L^1(\mathbb T)}<+\infty.
\end{eqnarray*}

 Since in addition 
 $
f(k)\mapsto -(\hat\beta(k)/4)f(k)
$
is bounded on $H^{-m}(\bbT)$,
the equation
\begin{eqnarray}
\label{limit-eqt}
d \bar\psi_*(t,k)&=&
 -\frac{\hat\beta(k)}{4} \bar \psi_*(t,k)dt +{\cal R}^{1/2}(t,k)dW(t,k),\nonumber\\
 \bar \psi_*(0,k)&=& \hat\psi(k)
 \end{eqnarray}
 has a unique $H^{-m}(\bbT)$-valued mild solution, by virtue of Theorem 7.4, p. 186 of \cite{DZ}.
It is given by the formula
 $$
 \bar\psi_*(t,k)=e^{-\hat \beta(k) t/4}\hat\psi+\int_0^te^{-\hat \beta(k) (t-s)/4}{\cal R}^{1/2}(s,k)dW(s,k).
 $$ 
We denote by  $H^{-m}_w(\bbT)$ the Hilbert space  equipped with the weak topology.
 Our main result is as follows.
\begin{thm}
\label{main-thm2}
Suppose that $3/2>m>1/2$ and both \eqref{011210a} and condition $\om)$ hold.
Then,  under the above assumptions,   the processes $\{\tilde \psi^{(\eps)}(t),\,t\ge0\}$ converge in law over 
$C([0,+\infty),H^{-m}_w(\bbT))$, as $\eps\to0+$,  to  $\{\bar\psi_*(t),\,t\ge0\}$.
\end{thm}
{\bf Remark.} As in the remark made after Theorem \ref{main-thm1} we can also conclude that
\begin{equation}
\label{040510b}
\lim_{t\to +\infty}\bbE\left|\langle\bar \psi_*(t)-\bar \psi_{s}(t),f\rangle\right|^2=0,
\end{equation}
where $ \bar\psi_s(t)$ is a  time homogeneous, distribution valued Ornstein-Uhlenbeck process given by
\begin{eqnarray}
\label{limit-eqt1b}
d \bar\psi_s(t,k)&=&
 -\frac{\hat\beta(k)}{4} \bar \psi_s(t,k)dt +\sqrt{\frac{\hat\beta(k) T}{2}}dW(t,k),\nonumber\\
 \bar \psi_s(0,k)&=& \hat\psi(k),
 \end{eqnarray}
 where $T=\|{\cal E}_0\|_{L^1(\bbT)}$.

\section{Proof of Theorem \ref{main-thm1}}
\label{sec4}

 The fact that the solution of \eqref{mollified-eqt} lies in $C(\bbT)$
 for each $\eps>0$ is a direct consequence of  Proposition
 \ref{prop010910} and  the embedding of $H^m(\bbT)$ into $C(\bbT)$ for
 $m>1/2$. 
We prove first the part $(i)$ of the theorem.
To explain the idea of the proof assume that  $n=1$ (that is, the
process $\hat\psi(t,k)$ for a fixed $k$),  
the independence of the compensated wave function for various $k$ 
is handled in the same manner. Since the  coefficients appearing in
the stochastic differential equation describing the  evolution of   
$\tilde \psi^{(\eps)}(t)$ (see \eqref{mollified-eqt}) are of the order
$O(1)$, it is easy to conclude that for  each $k$  the laws of the 
processes $\{\tilde \psi^{(\eps)}(t,k),\,t\ge0\}$ are tight over
$C([0,+\infty);\mathbb C)$, as $\eps\to0+$.  In order to identify the limit,  
thus  proving part i) of the theorem, we have to deal with the
rapidly oscillating terms. First, we show that  the rapidly
oscillating part of the bounded variation term in
\eqref{mollified-eqt}  
(with the factor $\exp\{2i\omega(k)t/\eps\}$ in (\ref{012808}))
vanishes in the limit thanks to part i) of  Corollary~\ref{cor2} below.


Next, the limit of the martingale part $\tilde {\cal M}_t^{(\eps)}(k)$ in (\ref{mollified-eqt}) 
is a complex Gaussian martingale with the quadratic variation equal to
$\int_0^t{\cal R}(s,k)ds$ thanks to the following:
\begin{equation}\label{mar0802}
\lim_{\eps\to0+}\sup_{t\in[0,t_*]}\left|\langle\tilde {\cal M}^{(\eps)}(k),(\tilde {\cal M}^{(\eps)})^*(k)\rangle_t-\int_0^t{\cal R}(s,k)ds\right|=0,
\end{equation}
where the convergence holds in probability, for any $t_*>0$.
This is  done in  Proposition~\ref{lm013108}. The method of proof of
(\ref{mar0802}) is as follows. 
From \eqref{060410}, we compute the quadratic variation:
\begin{eqnarray}
\label{050410}
&&
\langle \tilde{\cal M}^{(\eps)}(k),(\tilde {\cal M}^{(\eps)})^*(k)\rangle_t
\\
&&
=\sum_{\si_1,\si_2=\pm1}\si_1\si_2\int_0^tds\int_{\bbT}r^2(k,k')\hat\psi^{(\eps)}_{\si_1}(s,k-k')(\hat\psi^{(\eps)}_{\si_2})^*(s,k-k')\nonumber
 dk'.
\end{eqnarray} 
The terms appearing in \eqref{050410} are of the following form:
\begin{eqnarray}
&&{\cal V}_\eps^{(0)}(t):=\int_0^t\langle|\hat\psi^{(\eps)}(s)|^2,f\rangle ds,\nonumber\\
\label{mar804}&&
{\cal V}_{\eps}^{(1)}(t):=\int_0^t
\int_{\bbT}\hat\psi^{(\eps)}(s,k)\hat\psi^{(\eps)}(s,-k)f^*(k)dk \; ds.
\end{eqnarray}
Here $f(k)$ is a certain explicit function related to the scattering kernel. As $\hat\psi^{(\eps)}(t,k)$ (without the compensation) 
is rapidly oscillating as $e^{-i\omega(k)t/\eps}$, therefore we expect that only ${\cal V}_\eps^{(0)}(t)$ has a non-trivial limit.
This term contains no oscillation and is essentially the time integral of scattered energy $|\hat \psi^{(\eps)}(t,k)|^2$. It
has been shown in \cite{BOS} that the expectation of the energy converges to the solution of  \eqref{eq:phbolhom}. We need
to strengthen this result   to convergence in probability. 

The proof of part ii) of the theorem uses the same ideas. Integrating against a test function results in  the formula for the quadratic variation, see \eqref{080510a}, containing only  terms with fast oscillating factors, so the stochastic part vanishes  in the limit.

We now turn to the proof of part (i) the theorem. In particular,  we assume that $\hat\psi\in H^m$, $m>1/2$ so that
$\tilde\psi^{(\eps)}(t,k)$ is continuous and point-wise evaluations in $k$ make sense.
An application of the It\^o formula to \eqref{basic:sde:2} yields, see  Theorem 4.17 of \cite{DZ}, 
\begin{equation}
\label{wigner-eqt}
d|\hat\psi^{(\eps)}(t,k)|^2=\left[I_\eps(t,k)+I\!I_\eps(t,k)\right]dt
+d{\cal M}^{(\eps)}_t(k) + d {\cal M}^{(\eps)*}_t(k) ,
\end{equation}
where
\begin{eqnarray*}
&&I_\eps(t,k):=
(A[\hat\psi^{(\eps)}(t)])^*\left(k\right)\hat \psi^{(\eps)}\left(t,k\right)\vphantom{\int_0^1}+(\hat\psi^{(\eps)})^*\left(t,k\right)A[\hat \psi^{(\eps)}(t)]\left(k\right),
\\
&&
I\!I_\eps(t,k):=\sum_{y\in\bbZ}\left|Q[\hat\psi^{(\eps)}(t)](e_y)\left(k\right)\right|^2,
\end{eqnarray*}
 and $ {\cal M}^{(\eps)}_t $ is an ${\cal F}_t$-adapted local martingale, 
 given by
\begin{eqnarray*}
&&{\cal M}^{(\eps)}_t(k) =\int_0^t\hat \psi^{(\eps)}\left(s,k\right) (Q[\hat\psi^{(\eps)}(s)]
dW(s))^*\left(k\right) .
\end{eqnarray*}
From \eqref{040607} we obtain that
\begin{equation*}
I_\eps(t,k)=-\frac{\hat\beta(k)}{2}|\hat\psi^{(\eps)}(t,k)|^2-\frac{\hat\beta(k)}{4}\hat\psi^{(\eps)}_2(t,k),
\end{equation*}
where
$$
\hat\psi^{(\eps)}_2(t,k):=\hat\psi^{(\eps)}(t,k)\hat\psi^{(\eps)}(t,-k)+(\hat\psi^{(\eps)})^*(t,k)(\hat\psi^{(\eps)})^*(t,-k),
$$
while equation \eqref{053009}   yields
\begin{equation*}
I\!I_\eps(t,k)=\int_{\bbT}R(k,k')|\hat\psi^{(\eps)}(t,k')|^2dk'+\frac12\int_{\bbT}R(k,k')\hat\psi^{(\eps)}_2(t,k')dk'.
\end{equation*}

Analogous equation can be derived for
 $d[\hat\psi^{(\eps)}(t,k)\hat\psi^{(\eps)}(t,-k)]$. The corresponding
 terms shall be denoted by $\tilde I_\eps(t,k)$, $\tilde{I\!I}_\eps(t,k)$ and the martingale  
 $
 {\cal N}^{(\eps,1)}_{t}(k) +{\cal N}^{(\eps,2)}_{t}(k),
 $
 where
$$
\tilde I_\eps(t,k)=-\frac{2i\om(k)}{\eps}\hat\psi^{(\eps)}_2(t,k)+{\cal P}[\hat\psi^{(\eps)}(t),(\hat\psi^{(\eps)})^*(t)],
$$
\begin{equation}
\label{030207ba}
I\!I_\eps(t,k)={\cal Q}[\hat\psi^{(\eps)}(t),(\hat\psi^{(\eps)})^*(t)],
\end{equation}
where ${\cal P},{\cal Q}$ are second degree polynomials in $\hat\psi^{(\eps)}(t),(\hat\psi^{(\eps)})^*(t)$, and
\begin{eqnarray*}
&&{\cal N}^{(\eps,1)}_t(k) =\int_0^t\hat \psi^{(\eps)}\left(s,- k\right) (Q[\hat\psi^{(\eps)}(s)]
dW(s))\left(k\right) ,\nonumber\\
&&
\\
&&
{\cal N}^{(\eps,2)}_t(k) =\int_0^t\hat \psi^{(\eps)}\left(s,k\right)
(Q[(\hat\psi^{(\eps)}_{-1})^*(s)] dW(s))\left(-k\right).\nonumber
\end{eqnarray*}
\begin{prop}
\label{lm013108}
Let $f\in L^\infty(\mathbb T)$,  ${\cal V}_\eps^{(0)}(t)$ given by
\eqref{mar804}, and let ${\cal V}_{\eps,a}^{(1)}(t)$ be defined by
\begin{equation}
  \label{eq:1}
  {\cal V}_{\eps,a}^{(1)}(t):=\int_0^t
\int_{\bbT} \exp\left\{\frac{isa}{\eps}
\right\}\hat\psi^{(\eps)}(s,k)\hat\psi^{(\eps)}(s,-k)f^*(k)dk \; ds, \quad
a\in \bbR.
\end{equation}
 Then, 
for any $t_*>0$  we have
\begin{equation}
\label{013009}
\lim_{\eps\to0+}\sup_{t\in[0,t_*]}\left|{\cal V}_\eps^{(0)}(t)-\int_0^t\langle\bar{\cal E}(s),f\rangle ds\right|= 0
\end{equation}
and
\begin{equation}
\label{023009}
\lim_{\eps\to0+}\sup_{t\in[0,t_*]}\left|{\cal V}_{\eps,a}^{(1)}(t)\right|= 0,\quad a\in\bbR,
\end{equation}
in probability.
\end{prop}

The proof of this proposition shall be obtained at the end of  a series of lemmas.  
\begin{lm}
\label{lm023108}
For any $p\in[2,+\infty)$  there exists $C>0$ such that, for any $t_*>0$
\begin{equation}
\label{110510}
\sup_{\eps\in(0,1]}\bbE\left[\sup_{t\in[0,t_*]}
  \|\hat\psi^{(\eps)}(t)\|_{L^p(\bbT)}^p\right]\le 
Ce^{Ct_*}\|\hat\psi\|_{L^p(\bbT)}^p ,
\end{equation}
and, 
\begin{equation}
\sup_{\eps\in(0,1],k\in\bbT}\bbE\left[\sup_{t\in[0,t_*]}|\hat\psi^{(\eps)}(t,k)|^p\right]\le Ce^{Ct_*}\|\hat\psi\|^p_{L^p(\bbT)}.
\label{011310z}
\end{equation}
\end{lm}

{\bf Proof.} 
Let
$$
T_t^{(\eps)}\hat\psi(k):=\exp\left\{-i\frac{\om(k)t}{\eps}\right\}\hat\psi(k),\quad \hat \psi\in L^p(\bbT),\,t\in\bbR.
$$
We obviously have
\begin{equation}
\label{013108}
\|T_t^{(\eps)}\hat\psi\|_{L^p(\bbT)}= \|\hat\psi\|_{L^p(\bbT)},\quad\forall\,t\ge 0. 
\end{equation}
Using the Duhamel formula, the solution of \eqref{basic:sde:2} can be written as
\begin{eqnarray}\label{023108}
\hat\psi^{(\eps)}(t,k)=\hat\psi(k)+\int_0^tT_{t-s}^{(\eps)}B[\hat\psi^{(\eps)}(s)](k)ds 
+\int_0^tT_{t-s}^{(\eps)}Q[\hat\psi^{(\eps)}(s)]dW(s,k),
\end{eqnarray}
where $B f (k) = -\hat \beta(k) [f(k) - f^*(-k)]/4$.
 Hence, for a given $\eps\in(0,1]$ and $t_0>0$ to be adjusted later on, we can write
 \begin{eqnarray}
 \label{010410}
&&\bbE\left[\sup_{t\in[0,t_0]} | \hat\psi^{(\eps)}(t,k)|^p\right]\le C\left\{|\hat\psi(k)|^p+t_0^{p-1}\int_0^{t_0}\bbE |\hat\psi^{(\eps)}(s,k)|^pds\right.\nonumber\\
 &&
 +\left.\bbE\left\{\sup_{t\in[0,t_0]}\left|\int_0^tT_{-s}^{(\eps)}Q[\hat\psi^{(\eps)}(s)]dW(s,k)\right|^p\right\}\right\}.
 \end{eqnarray}
 To estimate the martingale term on the right hand side we use Burkholder-Davis-Gundy inequality which allows to bound it by
  \begin{eqnarray}
\label{020410}
4^{p/2}\bbE\left(\int_0^{t_0}\int_{\bbT}R(k,k')|\hat\psi^{(\eps)}(s,k-k')|^2dk'ds\right)^{p/2}
\le C_1t^{p/2-1}_0\int_0^{t_0}\bbE\|\hat\psi^{(\eps)}(s)\|^p_{L^p(\bbT)}ds,
 \end{eqnarray}
 for some constant $C_1>0$.
 Choosing $t_0$ sufficiently small, so that $Ct_0^p+CC_1t_0^{p/2}<1/2$, we conclude that
 \begin{equation}
 \label{011411}
\bbE\left\{\sup_{t\in[0,t_0]} \|\hat\psi^{(\eps)}(t)\|^p_{L^p(\bbT)}\right\}\le 2C\|\hat\psi\|^p_{L^p(\bbT)}.
 \end{equation}
The argument leading to \eqref{011411} can be used on each of the intervals $[jt_0, (j+1) t_0)$ for any $j\ge1$ and yields
 \begin{equation}
 \label{011411j}
 \bbE\left\{\sup_{t\in[jt_0,(j+1)t_0]} \|\hat\psi^{(\eps)}(t)\|^p_{L^p(\bbT)}\right\}\le C\bbE\|\hat\psi^{(\eps)}(jt_0)\|^p_{L^p(\bbT)}
\le C\bbE\left\{\sup_{t\in[(j-1)t_0,jt_0]} \|\hat\psi^{(\eps)}(t)\|^p_{L^p(\bbT)}\right\}, 
 \end{equation}
 for some constant $C>0$ independent of $j$ and $\eps\in(0,1]$.
Hence,
after $j$  iterations of the above estimate, we conclude 
  \begin{equation}
 \label{011411ja}
\bbE\left\{\sup_{t\in[jt_0,(j+1)t_0]} \|\hat\psi^{(\eps)}(t)\|^p_{L^p(\bbT)}\right\}\le C^j\|\hat\psi\|^p_{L^p(\bbT)}
 \end{equation}
 and \eqref{110510} follows.
Combining the above result with estimates \eqref{010410} and
\eqref{020410} we conclude  estimate \eqref{011310z}. \qed

Using 
the above lemma we conclude the following.
\begin{cor}
\label{cor2}
For  given $t_*>0$  and  function $f\in C^1[0,t_*]$ we have the following:
\begin{itemize}
\item[i)]
if $k\in\bbT$ and $a\in\bbR$ are such such that $-a\not=\om(k)$ then,
\begin{equation}
\label{070410}
\lim_{\eps\to0+}\bbE\left|\sup_{t\in[0,t_*]}\int_0^t\exp\left\{-i\frac{as}{\eps}\right\}f(s)\hat\psi_\eps(s,k)ds\right|=0,
\end{equation}
\item[ii)]
if $k,k'\in\bbT$ and $a\in\bbR$ are such that $-a\not=\om(k)+\om(k')$ then,
\begin{equation}
\label{030410}
\lim_{\eps\to0+}\bbE\left\{\sup_{t\in[0,t_*]}\left|\int_0^t\exp\left\{-i\frac{as}{\eps}\right\}f(s)\hat\psi^{(\eps)}(s,k)\hat\psi^{(\eps)}(s,k')ds\right|\right\}=0,
\end{equation}
\item[iii)] if $\om(k)+a\not=\om(k')$ then,
\begin{equation}
\label{030410a}
\lim_{\eps\to0+}\bbE\left\{\sup_{t\in[0,t_*]}\left|\int_0^t\exp\left\{-i\frac{as}{\eps}\right\}f(s)\hat\psi^{(\eps)}(s,k)(\hat\psi^{(\eps)})^*(s,k')ds\right|\right\}=0.
\end{equation}
\end{itemize}
\end{cor}{\bf Proof.} 
Using \eqref{basic:sde:2} we obtain  
\begin{eqnarray}
\label{090510}
&&
\exp\left\{-i\frac{at}{\eps}\right\}f(t)
\hat\psi_\eps(t,k)-f(0)
\hat\psi(k)
=-i\frac{a+\om(k)}{\eps}\int_0^t\exp\left\{-i\frac{sa}{\eps}\right\}f(s)\hat\psi_\eps(s,k)ds \nonumber
\\
&&
+\int_0^t{\cal P}[\hat\psi_\eps(s),(\hat\psi_\eps)^*(s)](k)ds+\int_0^t\sum_{y\in\bbZ}{\cal Q}_y[\hat\psi_\eps(s),(\hat\psi_\eps)^*(s)](k)w_y(ds),\end{eqnarray}
where ${\cal P}$, ${\cal Q}_y$ are first degree polynomials in
$\hat\psi_\eps(s)$, $(\hat\psi_\eps)^*(s)$ with bounded
coefficients. Using Lemma  \ref{lm023108} we have
$$
\bbE\left[\sup_{s\in[0,t_*]}\sum_{y\in\bbZ}|{\cal
    Q}_y[\hat\psi_\eps(s),(\hat\psi_\eps)^*(s)](k)|^2\right] \le
C\|\hat\psi\|_{L^2(\bbT)}^2. 
$$
Dividing both sides of \eqref{090510} by $(\om(k)+a)/\eps$ (possible since this factor is not equal to $0$)  we calculate
$$
\int_0^t\exp\left\{-i\frac{sa}{\eps}\right\}f(s)\hat\psi_\eps(s,k)ds.
$$
Using Lemma \ref{lm023108} we can easily conclude \eqref{070410}.

 The proofs of \eqref{030410} and \eqref{030410a} are analogous. We use the
 It\^o formula to express $d[\hat\psi^{(\eps)}(s,k)\hat\psi^{(\eps)}(s,k')]$ and $d[\hat\psi^{(\eps)}(s,k)(\hat\psi^{(\eps)})^*(s,k')]$. Then, we repeat the argument used above.
\qed

The following lemma shall be crucial for us.
\begin{lm}
\label{lm013009}
For  any $f\in L^2(\bbT)$, $t_*>0$  we have
\begin{equation}
\label{010310}
\lim_{\eps\to0+}\bbE\left[\sup_{t\in[0,t_*]}\left|\langle{\cal M}^{(\eps)}_t,f\rangle\right|^2\right]=0
\end{equation}
and
\begin{equation}
\label{020310}
\lim_{\eps\to0+}\bbE\left[\sup_{t\in[0,t_*]}\left|\langle{\cal N}^{(i,\eps)}_{t},f\rangle\right|^2\right]=0, \quad i=1,2.
\end{equation}
\end{lm}
{\bf Proof.} We only prove \eqref{010310}, the argument for \eqref{020310} is very similar.
We  write
\begin{eqnarray}
\label{030207a}
&&
\bbE\left|\langle{\cal M}^{(\eps)}_t,f\rangle\right|^2
\le2\left\{ \sum_{y\in\bbZ}\int_0^tds\,\bbE\left|\int_{\bbT^2} r(k,k')f^*(k)(\hat\psi^{(\eps)})^*(s,k-k')e_y^*(k')\hat\psi^{(\eps)}(s,k)d\bk\,\right|^2\right. \nonumber\\
&&\left. +\sum_{y\in\bbZ}\int_0^tds\bbE\left|\int_{\bbT^2} r(k,k')f^*(k)(\hat\psi^{(\eps)})(s,k'-k)e_y^*(k')\hat\psi^{(\eps)}(s,k)d\bk\right|^2\right\} .
\end{eqnarray}
Here, for abbreviation sake, we wrote $d\bk=dkdk'$.
 Using the Parseval identity we can further transform the right hand side of \eqref{030207a} into
 \begin{eqnarray}
\label{030207aa}
&&
2 \int_0^tds\int_{\bbT^3} r(k,k')r(k_1,k')f^*(k)f(k_1)\nonumber\\
&&\times
\left\{\bbE\left[(\hat\psi^{(\eps)})^*(s,k-k')\hat\psi^{(\eps)}(s,k)\hat\psi^{(\eps)}(s,k_1-k')(\hat\psi^{(\eps)})^*(s,k_1)\right]\right.\nonumber\\
&&\left. +\bbE\left[\hat\psi^{(\eps)}(s,k-k')\hat\psi^{(\eps)}(s,k)(\hat\psi^{(\eps)})^*(s,k_1-k')(\hat\psi^{(\eps)})^*(s,k_1)\right]\right\}d\bk,
\end{eqnarray}
 where $d\bk=dkdk_1dk'$.

Consider the term of  \eqref{030207aa} corresponding to the first expectation (the other can be dealt with in a similar fashion). 
Let
$$
{\cal K}_{1}= [(k,k',k_1):\om(k)+\om(k'-k_1)=\om(k')+\om(k-k_1)].
 $$
 Thanks to condition $\om)$  the three dimensional Lebesgue measure on $\bbT^3$  of the set vanishes.
 We claim that for $\bk=(k,k',k_1)\not\in {\cal K}_{1}$ we have
\begin{equation}
 \label{phi}
\lim_{\eps\to0+}\int_0^t\Psi^{(\eps)}(s,\bk)ds=0,
\end{equation}
where
$$
\Psi^{(\eps)}(s,\bk):=\bbE\left[(\hat\psi^{(\eps)})^*(s,k-k')\hat\psi^{(\eps)}(s,k_1-k')\hat\psi^{(\eps)}(s,k)(\hat\psi^{(\eps)})^*(s,k_1)\right].
 $$
 Using \eqref{basic:sde:2} and It\^o formula we conclude that
\begin{eqnarray}
 \label{010110}
&&
\frac{i}{\eps}\left[\om(k-k')+\om(k_1)-\om(k_1-k')-\om(k)\right] \int_0^t\Psi^{(\eps)}(s,\bk)ds\nonumber\\
&&
=\Psi^{(\eps)}(t,\bk)-\Psi^{(\eps)}(0,\bk)+\int_0^t{\cal P}[\hat\psi^{(\eps)}(s),(\hat\psi^{(\eps)})^*(s)](\bk)ds,
\end{eqnarray}
where ${\cal P}$ is a fourth degree polynomial formed over the wave function $\hat\psi^{(\eps)}(s)$, $(\hat\psi^{(\eps)})^*(s)$. 
Dividing both sides of \eqref{010110} by the factor in front of the integral on the left hand side and subsequently using 
 \eqref{011310z} with $p=4$ we conclude \eqref{phi}. The lemma then follows, provided we can substantiate the 
following interchange of the limit with  integral
\begin{eqnarray*}
 &&
 \lim_{\eps\to0+}\int_0^tds\int_{\bbT^3}r(k,k')r(k_1,k')f^*(k)f(k_1)\Psi^{(\eps)}(s,\bk)d\bk \\
&&
=\int_{\bbT^3}r(k,k')r(k_1,k')f^*(k)f(k_1)d\bk\left\{ \lim_{\eps\to0+}\int_0^t\Psi^{(\eps)}(s,\bk)ds\right\}.
\end{eqnarray*}
The latter however is a consequence of the Lebesgue dominated convergence theorem and \eqref{011310z}. 
This ends the proof of \eqref{010310}.
The proof of \eqref{020310} is analogous.
\qed


\subsection*{Proof of Proposition \ref{lm013108}  }
We first demonstrate  \eqref{023009}. It is a consequence of   parts ii) and iii) of Corollary \ref{cor2}, and the  Lebesgue dominated convergence theorem.  Indeed,
$$
\bbE\left\{\sup_{t\in[0,t_*]}\left|{\cal V}_{\eps,a}^{(1)}(t)\right|\right\}\le \bbE\left\{\int_{\bbT}dk\sup_{t\in[0,t_*]}\left|\int_0^t\exp\left\{\frac{isa}{\eps}\right\}\hat\psi^{(\eps)}(s,k)\hat\psi^{(\eps)}(s,-k)f^*(k)ds\right|\right\}. 
$$
Using condition $\om$) we conclude that the expression under the integral over $k$ on the right hand side vanishes, as $\eps\to0+$, possibly outside a set of $k$-s of null Lebesgue measure. Invoking again \eqref{011310z} we 
can substantiate exchanging of taking the limit and integration 
and \eqref{023009} follows.

As for  \eqref{013009}, observe that from the It\^o formula for 
 $d|\hat\psi^{(\eps)}(t,k)|^2$ we have
  \begin{eqnarray*}
 \langle|\hat\psi^{(\eps)}(t)|^2,f\rangle- \langle|\hat\psi(0)|^2,f\rangle =\int_0^t\langle{\cal L} |\hat\psi^{(\eps)}(s)|^2,f\rangle ds
+\frac12\int_0^t\langle {\cal L}\hat\psi^{(\eps)}_2(s),f\rangle ds+\langle{\cal M}^{(\eps)}_t,f\rangle.
\end{eqnarray*}
Denote by  $\{Q_{\eps}, \,\eps\in(0,1]\}$ 
the family of the laws of $\{|\hat\psi^{(\eps)}(t)|^2,\,t\ge0\}$  over $C([0,+\infty),L^2_w(\bbT))$. Here $L^2_w(\bbT)$ stands for the space $L^2(\bbT)$ equipped with the weak topology.

Using  Lemma \ref{lm023108}  we conclude from the above equality that for any $t_*>0$ there exists a constant $C>0$ such that
$$
\bbE \left|\langle|\hat\psi^{(\eps)}(t)|^2,f\rangle-\langle|\hat\psi^{(\eps)}(s)|^2,f\rangle\right|^4\le C(t-s)^2 ,\quad\forall\,\eps\in(0,1],\,t,s\in[0,t_*].
$$
This, according to   Theorem 12. 3 of  \cite{bil}, implies tightness of the family of the laws of $\{\langle|\hat\psi^{(\eps)}(t)|^2,f\rangle,\,t\ge0\}$, as $\eps\to0+$, over $C[0,+\infty)$ equipped with the usual topology of uniform convergence on compact intervals. From the above and  estimate \eqref{110510} we conclude weak pre-compactness of  $Q_{\eps}$, $\eps\in(0,1]$, see Theorem~3.1, p.  276 of \cite{jakubowski}.
Thanks to  Lemma \ref{lm013009} and the already proved formula \eqref{023009} we conclude that the limiting law is a $\delta$-type measure  supported on $\bar{\cal E}(t)$ --
 the solution of \eqref{100510}. This, in particular, implies that
 $$
\lim_{\eps\to0+} \sup_{t\in[0,t_*]}\left|\langle|\hat\psi^{(\eps)}(t)|^2-\bar{\cal E}(t),f\rangle\right|=0
 $$
 in probability. Hence \eqref{013009} follows.
%
\qed

\subsection*{Proof of part (i) of Theorem \ref{main-thm1}}

With the results proved above in hand, we return  to the proof of  part (i) Theorem  \ref{main-thm1}.
Assume first that $n=1$ and we consider the process $\tilde \psi^{(\eps)}(t,k)$ evaluated at a single $k$. 
From \eqref{mollified-eqt} and  \eqref{011310z} we conclude easily 
that for  any $t_*>0$ there exists a constant $C>0$ such that
$$
\bbE|\tilde \psi^{(\eps)}(t,k)-\tilde \psi^{(\eps)}(s,k)|^4\le C(t-s)^2,\quad\forall\,\eps\in(0,1],\,s,t\in[0,t_*].
$$
This implies tightness of  the laws of  $\{\tilde \psi^{(\eps)}(t,k),\,t\ge0\}$  over $C[0,+\infty)$.

In the next step we identify the limiting law
$P_k$ of $\{\tilde \psi^{(\eps)}(t,k),\,t\ge0\}$  over $C[0,+\infty)$. Denote by  $\Pi_t(f):=f(t)$, $f\in C[0,+\infty)$  the canonical coordinate map.

Consider the complex valued martingale
given by \eqref{060410}. Its quadratic variation is given by \eqref{050410}
 and, of course,
$\langle \tilde{\cal M}^{(\eps)}(k),\tilde {\cal M}^{(\eps)}(k)\rangle_t=0.$
Using Proposition \ref{lm013108} we conclude that
$$
\lim_{\eps\to0+}\sup_{t\in[0,t_*]}\left|\langle \tilde{\cal M}^{(\eps)}(k),(\tilde {\cal M}^{(\eps)})^*(k)\rangle_t-\int_0^t{\cal R}(s,k)ds\right|=0.
$$
Then by virtue of Theorem 5.4 of \cite{helland} we conclude that $\{ \tilde{\cal M}^{(\eps)}_t,\,t\ge0\}$ converge in law over $C[0,+\infty)$ to a complex valued Gaussian process  $\{ \tilde{\cal M}_t,\,t\ge0\}$ given by
\begin{equation}
\label{010510}
 \tilde{\cal M}_t(k):=\int_0^t{\cal R}^{1/2}(s,k) w(ds),
\end{equation}
where $\{ w(t),\,t\ge0\}$ is a complex valued standard Brownian motion. 

Assume now that
 $k\not=0$ and  $P_k$ is a limiting law of  $\{\tilde\psi^{(\eps)}(t,k),\,t\ge0\}$ obtained from a certain sequence $\eps_n\to0+$. Denote by $\Pi_t$ the coordinate mapping, given by $\Pi_t(g):=g(t)$ for $g\in C[0,+\infty)$. From \eqref{mollified-eqt} and \eqref{070410} we infer that 
 $$
 \Pi_t+\frac{\hat\beta(k)}{4}\int_0^t\Pi_sds,\quad t\ge0
 $$
is a $P_k$-martingale whose law coincides with that of the process described by \eqref{010510}.
The conclusion extends also to the case when $k=0$ and $\om(0)>0$. If, on the other hand, $\om(0)=0$ we have  $\hat \beta(0)=0$  and ${\cal R}^{1/2}(s,0)=0$ and  therefore 
$\Pi_t\equiv\Pi_0$ a.s.

 Suppose now that $k_1,\ldots,k_n\in \bbT$ are
pairwise distinct.  Denote by $Q_\eps$ the law of  
\[
\{(\tilde \psi^{(\eps)}(t,k_1),\ldots,\tilde \psi^{(\eps)}(t,k_n)),\,t\ge0\}
\]
over $C([0,+\infty),\mathbb C^n)$. Then,  we claim that
\begin{equation}
\label{101010}
\lim_{\eps\to0+}\sup_{t\in[0,t_*]}\left|\langle \tilde{\cal M}^{(\eps)}(k_i),(\tilde {\cal M}^{(\eps)})^*(k_j)\rangle_t-\delta_{i,j}\int_0^t{\cal R}(s,k_i)ds\right|=0
\end{equation}
and, obviously,
\begin{equation}
\label{111010}
\lim_{\eps\to0+}\sup_{t\in[0,t_*]}\left| \langle \tilde{\cal M}^{(\eps)}(k_i),\tilde {\cal M}^{(\eps)}(k_j)\rangle_t\right|=0,\quad \forall \,i,j=1,\ldots,n.
\end{equation}
To see \eqref{101010} note that for $i\not=j$ we have
\begin{eqnarray*}
&&
\langle \tilde{\cal M}^{(\eps)}(k_i),(\tilde {\cal M}^{(\eps)})^*(k_j)\rangle_t=\sum_{\si_1,\si_2=\pm1}\si_1\si_2\int_0^t\exp\left\{\frac{i(\om(k_i)-\om(k_j))s}{\eps}\right\}ds
\\
&&
\times\int_{\bbT}r(k_i,k')r(k_j,k')\hat\psi^{(\eps)}_{\si_1}(s,k_i-k')(\hat\psi^{(\eps)}_{\si_2})^*(s,k_j-k') dk',
\end{eqnarray*}
Using part iii) of Corollary \ref{cor2} combined  with condition
$\om)$ we conclude, thanks to the fact that $k_i\not=k_j$, that 
$$
\lim_{\eps\to0+}
\sup_{t\in[0,t_*]}\left|\int_0^t\exp\left\{\frac{i(\om(k_i)-\om(k_j))s}{\eps}\right\}
  \hat\psi^{(\eps)}_{\si_1}(s,k_i-k')(\hat\psi^{(\eps)}_{\si_2})^*(s,k_j-k')ds\right|=0
$$
for a.e. $k'\in\bbT$. Using  \eqref{011310z} in the same way as in the proof of \eqref{023009} we can substantiate exchanging the passage to the limit with the respective integration and conclude \eqref{101010}.

 Combining \eqref{101010} and \eqref{111010} with \eqref{070410} we obtain from equation \eqref{mollified-eqt} that any limiting point of the family of laws of $Q_{\eps_n}$ as $\eps_n\to0+$ is a measure $P_{k_1,\ldots,k_n}$ such that
$$
{\cal M}_t=({\cal M}_t^{(1)},\ldots,{\cal M}_t^{(n)}):= \Pi_t+\frac{\hat\beta(k)}{4}\int_0^t\Pi_sds,\quad t\ge0
 $$
is $\mathbb C^n$-valued martingale, whose quadratic  covariation is given by
$$
\langle {\cal M}^{(i)},({\cal M}^{(j)})^*\rangle_t=\delta_{i,j}\int_0^t{\cal R}(s,k_j)ds
$$
and
$$
\langle {\cal M}^{(i)},({\cal M}^{(j)})\rangle_t=0,\quad \forall \,i,j=1,\ldots,n.
$$
 This of course implies that $P_{k_1,\ldots,k_n}=P_{k_1}\otimes\ldots\otimes P_{k_n}$.

\subsection*{Proof of part ii) of Theorem \ref{main-thm1}}
\label{sec:proof-part-ii}

Let $f\in L^2(\bbT)$. We shall prove that 
\begin{eqnarray}
\label{050410a1}
&&
\lim_{\eps\to0+}\bbE|\langle \tilde{\cal M}^{(\eps)}_t,f\rangle|^2=0.
\end{eqnarray}
Assuming this result we show how to finish the proof of part (ii).  Denote 
$$
\delta \psi^{(\eps)}(t):= \tilde \psi^{(\eps)}(t)- \bar\psi(t).
$$ 
Using  Lemma \ref{lm023108} and  Theorem 3.1, p.  276 of \cite{jakubowski} 
we can  conclude weak pre-compactness of $P_{\eps}$, $\eps\in(0,1]$ --
the family of the laws of $\{\delta \psi^{(\eps)}(t),\,t\ge0\}$ -- in $C([0,+\infty),L^2_w(\bbT))$. 
With the help of Corollary \ref{030410} and \eqref{050410a1} we conclude that
the limiting measure, as $\eps\to0+$,   is supported on the solution of the equation
$$
\langle g(t),f\rangle-\frac{1}{4}\int_0^t\langle \hat \beta g(s),f\rangle ds=0,\quad\forall\,f\in L^2(\bbT).
$$
This of course shows that it is the $\delta$-measure supported on $g(t)\equiv0$. Hence, in particular we get
 \begin{equation}
 \label{060510}
\lim_{\eps\to0+}\sup_{t\in[0,t_*]}|\langle\delta \psi^{(\eps)}(t),f\rangle|=0
\end{equation}
in probability and   \eqref{040510} follows.

Coming back to the proof of  \eqref{050410a1} note that by the definition of the martingale  $\tilde{\cal M}^{(\eps)}_t$, see \eqref{060410}, we only need to show that
\begin{eqnarray}
\label{080510}
\lim_{\eps\to0+}\bbE\left|\int_0^t\int_{\bbT^2}\exp\left\{i s\frac{\om(k)}{\eps}\right\}r(k,k')f^*(k)\right.
\left.\vphantom{\int_0^1}\hat\psi^{(\eps)}_{\si}(s,k-k')dW(s,k')dk\right|^2=0
\end{eqnarray}
for $\si=\pm1$. We consider only the case $\si=1$, the other one can be dealt in a similar manner.
The expression under the limit in \eqref{080510} equals 
\begin{equation}
\label{080510a}
\int\limits_0^t\int\limits_{\bbT^3}\exp\left[i s\frac{\om(k)-\om(k_1)}{\eps}\right]\!
r(k,k')r(k_1,k')f^*(k)f(k_1)
\vphantom{\int_0^1}\bbE\left[\hat\psi^{(\eps)}(s,k-k')(\hat\psi^{(\eps)})^*(s,k_1-k')\right]dsd\bk,
\end{equation}
with $d\bk=dkdk_1dk'$. 
%
%
Using Corollary \ref{cor2} and an argument identical with the one used in the proof of Lemma \ref{lm013009} we conclude that 
\begin{eqnarray*}
\lim_{\eps\to0+}\int_0^t\exp\left\{i s\frac{\om(k)-\om(k_1)}{\eps}\right\} 
\bbE\left[\hat\psi^{(\eps)}(s,k-k')(\hat\psi^{(\eps)})^*(s,k_1-k')\right]ds=0
\end{eqnarray*}
for all $k',k,k_1$ such that $\om(k-k')+\om(k_1)-\om(k)\not=\om(k_1-k')$. Since the latter inequality holds
on the set of null Lebesgue measure we conclude
equality in \eqref{080510}, thanks to the Lebesgue dominated convergence theorem.

\section{Spatially homogeneous initial data}

\label{sec5}

 Tightness of the family of laws $\{\tilde\psi^{(\eps)}(t),\,t\ge0\}$, in the space  of continuous functionals taking values in a 
 space of distributions is again due to the fact that the  evolution equation \eqref{mollified-eqt} contains no terms that are large in magnitude.  
 This is done in Sections~\ref{sec5.1} and~\ref{sec5.2}.
 However, we have no estimates of the $H^{-m}(\bbT)$ norm of $\tilde\psi^{(\eps)}(t)$ analogous to the ones in 
 Lemma \ref{lm023108}, that have played an important role in the limit identification argument of 
 Section \ref{sec4} for square integrable data.  Therefore, instead of considering the quadratic variation of the martingale term as
 we did in the proof of Theorem~\ref{main-thm1}, for the proof of Theorem~\ref{main-thm2} we identify the limit of all moments 
of $\tilde\psi^{(\eps)}(t)$. Accordingly,   we first write equations for time evolution of an arbitrary moment of $\tilde\psi^{(\eps)}(t)$
in Section \ref{sec5.3}. Using standard averaging argument we show (see Proposition \ref{012410})  the convergence of moments, 
as $\eps\to0+$, to a solution of the limiting equation obtained simply
by discarding the oscillatory terms from the moment equation. Finally in Section \ref{sec5.5}
 we prove that the solutions of the limiting equation coincide with the respective moments of the non-homogeneous Ornstein-Uhlenbeck equation \eqref{limit-eqt} concluding in this way the proof of Theorem \ref{main-thm2}.

\subsection{Properties of spatially homogeneous solutions of \eqref{basic:sde:2}}

\label{sec5.1}
The initial data $\hat \psi$ considered in this section is  random and takes values in the Hilbert space of distributions $H^{-m}(\bbT)$ 
for some $m>1/2$. In fact, in Sections \ref{sec5.1}-\ref{sec5.4} we shall not make any use of the assumption that the data is 
Gaussian and we use only the fact that it is spatially homogeneous and
 \begin{equation}
 \label{031210}
 \bbE\|\hat\psi\|^2_{H^{-m}(\bbT)}<+\infty.
 \end{equation}
  Gaussianity shall be used only in Section \ref{sec5.5}.

Consider the random field $\{\psi_y:=\langle \hat\psi,e_y\rangle,\,y\in\bbZ\}$.  The field is assumed to be spatially homogeneous, i.e.   $\{\psi_{y+z},\,y\in\bbZ\}$ and $\{\psi_{y},\,y\in\bbZ\}$ have identical laws for all $z\in\bbZ$,  and centered, i.e. $\bbE\psi_0=0$. 
Spatial homogeneity is equivalent to the fact that $\hat \psi(k)$ and $e_z(k)\hat \psi(k)$ are identically distributed in $H^{-m}(\bbT)$ for any $z\in\bbZ$. 
Note that, since $m>1/2$, 
$$
\sum_{y\in\bbZ}(1+y^2)^{-m}\bbE|\psi_y|^2=\bbE\|\hat\psi\|^2_{H^{-m}(\bbT)}<+\infty,
$$
due to \eqref{031210}.


 Since the covariance function of the field
$$
S_{x-y}:=\bbE[\psi_x\psi^*_y],\quad\,\forall\, x,y\in\bbZ
$$
is positive definite, there exists a finite measure $\hat E(dk)$ such that
$$
S_x=\int_{\bbT}e^{i x k}\hat E(dk),\quad\forall\,x\in\bbZ.
$$ 
We assume that  the covariance function decays sufficiently fast in space so that
\begin{equation}
\label{conv-y}
\sum_{x\in\bbZ}(|\bbE[\psi_x^*\psi_0]|+|\bbE[\psi_x\psi_0]|)<+\infty.
\end{equation}
Assumption \eqref{conv-y} 
implies,   in particular,  that $\hat E(dk)={\cal E}_0(k)dk$ for some non-negative energy density ${\cal E}_0\in C(\bbT)$ and both this function
and 
$
{\cal Y}=\sum_{x\in\bbZ}e_x\bbE[\psi_x\psi_0]
$
belong to  $C(\bbT)$. 
When the field $\psi_x$ is a complex valued Gaussian, as described Section \ref{sec2.3.2}, we have ${\cal Y}\equiv0$.
This and \eqref{011210a} together imply \eqref{conv-y}.

We note that the translation invariance of the solution persists in time. Indeed, 
let $\psi^{(\eps)}_x(t):=\langle \hat\psi^{(\eps)}(t),e_x\rangle$ and $z\in\bbZ$. A direct computation shows that   $e_z\hat\psi^{(\eps)}(t)$ is 
also a solution of  \eqref{mollified-eqt}. Since 
the laws of  the initial conditions $e_z\hat\psi$ and that of  $\hat\psi$ are identical, we conclude from the uniqueness in law of 
solutions   that the same holds 
for the processes
$\{e_z\hat\psi^{(\eps)}(t),\,t\ge0\}$ and $\{\hat\psi^{(\eps)}(t),\,t\ge0\}$. 
In consequence, the laws of $\{\psi^{(\eps)}_x(t),\,x\in\bbZ\}$ and that of $\{\psi^{(\eps)}_{x+z}(t),\,x\in\bbZ\}$ 
are identical for any $z\in\bbZ$. 
We can now define the correlation functions
$$
S^{(\eps)}_{t,x}=\bbE\left[\psi^{(\eps)}_x(t)(\psi^{(\eps)}_0)^*(t)\right]\quad\mbox{and}\quad Y^{(\eps)}_{t,x}=\bbE\left[\psi^{(\eps)}_x(t)\psi^{(\eps)}_0(t)\right] 
$$
and introduce two distributions on $H^{-m}(\bbT)$
$$
\langle f, \hat S^{(\eps)}_{t}\rangle :=\sum_{x\in\bbZ} 
\check f_x(S^{(\eps)}_{t,x})^*\quad\mbox{and}\quad 
\langle f, \hat Y^{(\eps)}_{t}\rangle :=\sum_{x\in\bbZ} \check f_x(Y^{(\eps)}_{t,x})^*.
$$
We recall the following result of~\cite{BOS}.
\begin{prop}
\label{lm011310}
For any $\eps\in(0,1]$ and $t\ge0$  we have $\hat S^{(\eps)}_{t},\hat Y^{(\eps)}_{t}\in L^1(\bbT)$. Moreover,
\begin{itemize}
\item[(1)] $\hat S^{(\eps)}_{t}$ is non-negative, and for any $t_*>0$
\begin{equation}
\label{011310}
\sup_{\eps\in(0,1]}\sup_{t\in[0,t_*]}(\|\hat S^{(\eps)}_{t}\|_{L^1(\bbT)}+\|\hat Y^{(\eps)}_{t}\|_{L^1(\bbT)})<+\infty,
\end{equation}
\item[(2)]
for any $f\in L^\infty(\bbT)$ we have
\begin{equation}
\label{021310}
\lim_{\eps\to0+}\sup_{t\in[0,t_*]}\left|\langle\hat S^{(\eps)}_{t}-\bar{\cal E}(t),f\rangle\right|=0,
\end{equation}
where $\bar{\cal E}(t)$ is given by \eqref{100510} with the initial condition replaced by ${\cal E}_0(k)$ 
\item[(3)]
for any $f$ such that $f\om^{-1}\in L^\infty(\bbT)$ we have 
\begin{equation}
\label{021310a}
\lim_{\eps\to0+}\sup_{t\in[0,t_*]}\left|\int_0^t\langle\hat Y^{(\eps)}_{s},f\rangle ds\right|=0.
\end{equation}
\end{itemize}
\end{prop}
{\bf Proof.}  Parts 1) and  2) of the lemma are contained in Lemma 12 and  Theorem 10 of   \cite{BOS}, respectively. Part 3) follows easily from part 1) and the arguments used in the proof of Corollary \ref{cor2}. 
\qed
\subsection{Tightness of solutions of \eqref{mollified-eqt}}\label{sec5.2}

Given $f\in H^m(\bbT)$, we denote by $Q_\eps$ and $Q_{\eps,f}$ the laws of
the processes $\{\hat\psi^{(\eps)}(t),\,t\ge0\}$ and  $\{\langle f, \hat\psi^{(\eps)}(t)\rangle,\,t\ge0\}$ over $C([0,+\infty),H^{-m}_w(\bbT))$ and $C([0,+\infty),\mathbb C)$, respectively,
and by $\{\tilde Q_{\eps},\,\eps\in(0,1]\}$ the family of laws of $\{\tilde\psi^{(\eps)}(t),\,t\ge0\}$  over $C([0,+\infty),H^{-m}_w(\bbT))$.
According to \cite{mitoma}, see Remark R1, p. 997, to verify the tightness of  $\tilde Q_{\eps}$, 
it suffices to show the following two conditions:
\begin{itemize}
\item[(UC)]
for  any $\sigma,M,t_*>0$ there exists a $\delta>0$ such that
$$
\bbP\left[\sup_{t\in[0,t_*]}|\langle\tilde\psi^{(\eps)}(t),f\rangle|\ge M\right]<\si,\quad\forall\,\|f\|_{H^m(\bbT)}<\delta,\quad \eps\in(0,1],
$$
and
\item[(FDT)]
for  any $f\in H^{m}(\bbT)$  the family of the laws of the processes
$
\{\langle\tilde\psi^{(\eps)}(t),f\rangle,\,t\in[0,t_*]\}$, $\eps\in(0,1]
$
is tight over $C[0,t_*]$ for any $t_*>0$.
\end{itemize}
As in  \eqref{021310aa}  we conclude that 
 for any $f_1,f_2\in H^m(\bbT)$, where $m>1/2$, the covariance 
\begin{equation}
\label{031310}
\bbE\left[\langle f_1,\hat\psi^{(\eps)}_t\rangle\langle f_2,\hat\psi^{(\eps)}_t\rangle^* \right]=\int_{\bbT}\hat S_t^{(\eps)}(k)f_1(k)f_2^*(k)dk.
\end{equation}
From \eqref{mollified-eqt} and Doob's inequality there exists a constant $C>0$ such that
\begin{equation}
\label{041310}
\bbE\left[\sup_{t\in[0,t_*]}|\langle\tilde\psi^{(\eps)}(t),f\rangle|^2\right]\le C\left\{\bbE |\langle\hat\psi,f\rangle|^2\right.
\left.\!+\!\int_0^{t_*}
\bbE\left|\left\langle {\cal  A}\left[\frac{t}{\eps},\tilde\psi^{(\eps)}(t)\right],f\right\rangle\right|^2dt 
+\bbE\left|\left\langle\tilde{\cal M}^{(\eps)}_{t_*},f\right\rangle\right|^2\right\}. 
\end{equation}
Using \eqref{031310}, \eqref{011310} and the definitions of $ {\cal  A}[t/\eps,\cdot]$, and the martingale 
$\tilde{\cal M}^{(\eps)}_t$ (see  \eqref{012808} and \eqref{060410}) we conclude that the right hand side of \eqref{041310} 
can be estimated from above by $C\|f\|_\infty^2$, which can be made less than $\si>0$, provided we choose $\delta>0$ sufficiently 
small.

To show condition (FDT) consider  $\tilde Q^{(M)}_{\eps,f}$ -- the law of the stopped process
\[
\{(\langle\tilde\psi^{(\eps)}(t\wedge \tau_M^{(\eps)}),f\rangle,\langle\tilde\psi^{(\eps)}(t\wedge \tau_M^{(\eps)}), f_0\rangle)\,t\in[0,t_*]\} 
\]
over $C([0,t_*];\mathbb C^2)$. Here $ f_0(k):=f(-k)$ and
$$
\tau_M^{(\eps)}:=\inf[t\in[0,t_*]:|\langle\tilde\psi^{(\eps)}(t),f\rangle|^2+|\langle\tilde\psi^{(\eps)}(t),  f_0\rangle|^2\ge M^2].
$$
We adopt the convention that $\tau_M:=t_*$ if the set is empty. 
Thanks to (UC) we conclude that
$\lim_{M\to+\infty}\tau_M^{(\eps)}=t_*,
$ a.s. for each $\eps\in(0,1]$. 
Denote also by $\tilde Q_{\eps,f}$  the law of the  process without the stopping condition. 

From \eqref{mollified-eqt} we conclude that for a fixed $M$ and an arbitrary non-negative function 
$\phi:\mathbb C^2\to\bbR$, of class $C^1_c(\bbR^4),$ one can choose a constant $K_\phi$, 
independent of spatial translations of $\phi$, such that
$$
\phi(\langle\tilde\psi^{(\eps)}(t\wedge \tau_M^{(\eps)}),f\rangle,\langle\tilde\psi^{(\eps)}(t\wedge \tau_M^{(\eps)}), f_0\rangle)+K_\phi t,\,t\in[0,t_*]
$$
is a non-negative submartingale. This proves tightness of  $\{\tilde Q^{(M)}_{\eps,f},\,\eps\in(0,1]\}$ for a fixed $M$, by virtue of 
Theorem 1.4.3 of \cite{stroock-varadhan}. Since for any $\si>0$ one can find a sufficiently large $M>0$ such that  $B_M$ --
the ball centered at $0$ and of radius $M$ in $ C([0,t_*];\mathbb C^2)$ -- satisfies
$$
\tilde Q^{(M)}_{\eps,f}(B^c_M)+ \tilde Q_{\eps,f}(B^c_M)<\si
$$
and 
$$
\tilde Q^{(M)}_{\eps,f}(B_M\cap A)=\tilde Q_{\eps,f}(B_M\cap A)
$$
for all  Borel measurable subsets $A$  of  $ C([0,t_*];\mathbb C^2)$, we conclude  
tightness of $\{\tilde Q_{\eps,f},\,\eps\in(0,1]\}$,  see  
step (vi) of the proof of Theorem 3 of~\cite{kesten-papanicolaou} for details of this argument.

\subsection{Evolution of  moments}

\label{sec5.3}

To describe the evolution of moments  we rewrite equation
\eqref{mollified-eqt} in a more compact form, as a $2 \times2$  linear system of
equations with multiplicative noise.
Denote by ${\bf C}(t,k)=[C_{ij}(t,\bk)]$, $i,j=\pm1$, the $2\times 2$ hermitian  matrix
$$
{\bf C}(t,k):=\left[
\begin{array}{ll}
C_{1,1}&C_{1,-1}\\
C_{-1,1}&C_{-1,-1}
\end{array}\right],
$$
with the entries
$$
C_{p,q}(t,k):=\frac{pq\hat\beta(k)}{4}\exp\left\{ip\om(k)(1-pq)t\right\}.
$$
Let also ${\bf Q}(t,k,k')=[ Q_{pq}(t,k,k')]$, $p,q=\pm1$, be the $2\times 2$  matrix
$$
Q_{p,q}(t,k,k'):=ipqr(k,k-k')e^{ip[\om(k)-pq\om(k')]t}
$$
and
 $W(t,k):=\sum_ye_y(k)w_y(t)$. Let us recall that $\tilde\psi_{-1}^{(\eps)}(t,k)=\tilde\psi^{(\eps)*}(t,-k)$. Then, equation for 
$$
\Psi^{(\eps)}(t,k)=\left[
\begin{array}{c}
\tilde\psi^{(\eps)}(t,k)\\
\\
\tilde\psi^{(\eps)}_{-1}(t,k)
\end{array}\right]
$$
is
\begin{eqnarray}
\label{031710}
&&\!\!\!\!\!\!d\Psi^{(\eps)}(t,k)=-{\bf C}\left(\frac{t}{\eps},k\right)\Psi^{(\eps)}(t,k)dt
+\int_{\bbT}{\bf Q}\left(\frac{t}{\eps},k,k-k'\right)\Psi^{(\eps)}(t,k-k') W(dt,dk'),\nonumber\\
&&\!\!\!\!\!\!\Psi^{(\eps)}(0,k)=\Psi(k),
\end{eqnarray}
with the initial data
$$
\Psi(k)=\left[
\begin{array}{c}
\hat\psi(k)\\
\\
\hat\psi_{-1}(k)
\end{array}\right].
$$
Let $\{{\bf S}_\eps(s,t,k),\,s,t\in\bbR\}$ be the   $2\times 2$ Hermitian matrices solving the deterministic system
\begin{eqnarray*}
&&
\frac{d{\bf S}_\eps(s,t,k)}{dt}=-{\bf C}\left(\frac{t}{\eps},k\right){\bf S}_\eps(s,t,k)\\
&&
{\bf S}_\eps(s,s,k)=I_2.
\end{eqnarray*}
Here $I_2$ is the $2\times 2$ identity matrix.
Existence and uniqueness of solutions to \eqref{031710} in the strong sense (thus implying the result in the mild, or weak sense as well) follows from an
argument used in Chapter 6 of \cite{DZ} (because the  generators for the evolution family  ${\bf S}_\eps(s,t)$ are bounded),  
see Proposition 6.4 there. Although the case considered here differs slightly  
because the coefficients are time dependent, this does not influence the results.

Given a nonnegative integer $p\ge1$, define a  tensor valued distribution  on  $H^{-m/p}(\bbT^{p})$
$$
\hat M^{(\eps)}(t):=
\left[ \hat M_{\bi}^{(\eps)}(t)\right],\,\quad \bi=(i_1,\ldots,i_p)\in\{-1,1\}^p,
$$
by  
$$
\hat M_{\bi}^{(\eps)}(t) =\bbE\left[\tilde \psi^{(\eps)}_{i_1}(t)\otimes \ldots\otimes \tilde \psi^{(\eps)}_{i_p}(t)\right].
$$
Note that also
\begin{equation}
\label{021910}
\hat M_{\bi}^{(\eps)}(0) =\hat M_{\bi}:=\bbE\left[\hat \psi_{i_1}\otimes \ldots\otimes \hat \psi_{i_p}\right]
\end{equation}

For a given multi-index $\bi$ we define the multi-indices 
$
\bi_{\ell}(j)=(i_1',\ldots,i_p')$, $
\bi_{\ell,m}(j_1,j_2)=(i_1'',\ldots,i_p'')$ given by: $i_q'=i_q$ for $q\not=\ell$ and  $i_\ell'=j$, and
$i_q''=i_q$ for $q\not=\ell,m$ and  $i_\ell''=j_1$, $i_m''=j_2$.
Denote by ${\cal M}(\bbT^p)$ the space of all complex valued Borel measures $\nu$  on $\bbT^p$ whose  total variation norm $\|\nu\|_{\rm TV}$ is finite. 
\begin{prop}
\label{prop011810}
The following are true:
\begin{itemize}
\item[1)]
$\hat M^{(\eps)}(t)$ is the unique solution in $H^{-m/p}(\bbT^{p})$ of the system of equations
\begin{eqnarray}
\label{011910}
&&\frac{d}{dt}\hat M_{\bi}^{(\eps)}(t,\bk)=-\sum_{\ell=1}^p\sum_{j=\pm1}C_{i_\ell,j}\left(\frac{t}{\eps},k_\ell\right)\hat M_{\bi_\ell(j)}^{(\eps)}(t,\bk)\\
&&
+\sum_{1\le \ell<m\le p}\sum_{j_1,j_2=\pm1}\int_{\bbT}{\cal R}_{i_\ell,i_m}^{j_1,j_2}\left(\frac{t}{\eps},k_\ell,k_m,k'\right) \hat M_{\bi_{\ell,m}(j_1,j_2)}^{(\eps)}(t,\bk_{\ell,m}')dk',\nonumber
\end{eqnarray}
with $\bi\in\{-1,1\}^p$ and the initial data given by \eqref{021910}.
Here 
$$
{\cal R}_{i_\ell,i_m}^{j_1,j_2}\left(\frac{t}{\eps},k_\ell,k_m,k'\right):=Q_{i_\ell,j_1}\left(\frac{t}{\eps},k_\ell,k'_\ell\right)Q_{i_m,j_2}\left(\frac{t}{\eps},k_m,k'_m\right)
$$
and $\bk_{\ell,m}'=(k_1',\ldots,k_p')$, where $k_p':=k_p$ for $p\not=\ell,m$ and $k_\ell':=k_\ell-k'$, $k_m':=k_m+k'$.
\item[2)] If the initial condition is from  ${\cal M}(\bbT^p)$ then the solution also belongs to ${\cal M}(\bbT^p)$ and for any $t_*>0$ 
\begin{equation}
\label{062910}
M_*(T):=\sum_{\bi\in\{-1,1\}^p}\sup_{\eps\in(0,1]}\sup_{t\in[0,t_*]}\|\hat M_{\bi}^{(\eps)}(t)\|_{\rm TV}<+\infty.
\end{equation}
\end{itemize}
\end{prop}
{\bf Proof.} 
The fact that $\hat M^{(\eps)}(t)$ is a solution of \eqref{011910} follows by an application of It\^o formula and equation \eqref{031710}.
Since the operators appearing on the right hand side of the equation in question are uniformly Lipschitz, on any compact time interval, both in $H^{-m/p}(\bbT^p)$ and ${\cal M}(\bbT^p)$  the proof of uniqueness of solutions in these spaces  
 is standard. Estimate \eqref{062910} follows by an application of Gronwall's inequality.  
\qed

\subsection{Asymptotics of even moments}
\label{sec5.4}

Let us now describe the limit moment equations.
Assume that $p=2n$ is even, then
for any $1\le \ell<m\le 2n$ let 
$D_{\ell,m}:=[\bk\in\bbT^{2n}:k_\ell=-k_m]$. We define a bounded linear operator ${\cal R}_{\ell,m}:{\cal M}(\bbT^{2n})\to {\cal M}(\bbT^{2n})$ by
\begin{eqnarray*}
&&
\int_{\bbT^{2n}} fd {\cal R}_{\ell,m}\nu:=\int_{\bbT}dk\left\{\int_{D_{\ell,m}}r^2(k,k-k'_\ell)f(S(\bk',k))\nu(d\bk')\right\}
\end{eqnarray*}
 for any
bounded, measurable $f:\bbT^{2n}\to\mathbb C$ and $\nu\in  {\cal M}(\bbT^{2n})$.
 We define $S:\bbT^{2n+1}\to\bbT^{2n}$ as follows: given  $\bk'=(k_1',\ldots,k_{2n}')\in \bbT^{2n}$ and $k\in\bbT$ we let
  $(k_1,\ldots,k_{2n})=S(\bk',k)$ if  $k_j=k'_j$ for $j\not\in\{\ell,m\}$ and $k_\ell=k$, $k_m=-k$.

Suppose that  the components of the tensor $\hat M=[\hat M_{\bi}]$ belong to  ${\cal M}(\bbT^{2n})$. Similarly to part~1) 
of Proposition \ref{prop011810} we conclude that the initial value problem 
\begin{eqnarray}
&&\frac{d}{dt}\hat M_{\bi}(t)=-\frac{1}{4}\left(\sum_{\ell=1}^{2n}\hat\beta\left(k_\ell\right)\right)\hat M_{\bi}(t)+\sum_{1\le \ell<m\le 2n}\sum_{j=\pm1}{\cal R}_{\ell,m} \hat M_{\bi_{\ell,m}(j,-j)}(t),\nonumber
\label{031910}\\
&&
\hat M(0)=\hat M.
\end{eqnarray} 
possesses a unique solution in $C([0,+\infty),{\cal M}(\bbT^{2n}))$.

Any partition of the set $\{1,\ldots,2n\}$ into a disjoint set of pairs is called a pairing.
Define
$$
\mu(d\bk)=\sum_{\cal F}\prod_{(\ell,m)\in{\cal F}}\delta(k_\ell+k_m)d\bk,
$$
where $d\bk=dk_1\ldots dk_{2n}$ and the summation extends over all possible pairings of $\{1,\ldots,2n\}$. The measure is supported in 
$\bbH:=\bigcup_{\cal F}\bbH({\cal F})$ where 
$$
\bbH({\cal F}):=[\bk:k_\ell+k_m=0,\,\forall\,(\ell,m)\in{\cal F}].
$$
Suppose that the components of the tensor $\rho(\bk)=[\rho_{\bi}(\bk)]$, $\bi\in\{-1,1\}^{2n}$ belong to $L^1(\mu)$. Consider
the initial value problem 
\begin{eqnarray}
&&\frac{d}{dt}\rho_{\bi}(t,\bk)=-\frac{1}{4}\left(\sum_{\ell=1}^{2n}\hat\beta\left(k_\ell\right)\right)\rho_{\bi}(t,\bk)\nonumber\\
&&
+\sum_{1\le \ell<m\le 2n}\sum_{j=\pm1}\int_{\bbT} r^2(k_\ell, k_\ell-k')1_{D_{\ell,m}}(\bk)\rho_{\bi_{\ell,m}(j,-j)}(t,\bk'_{\ell,m})dk',\nonumber\\
&&
\rho_{\bi}(0,\bk)=\rho_{\bi}(\bk),\, \bi\in\{-1,1\}^{2n},
\label{031910a1}
\end{eqnarray}
with   $\bk'_{\ell,m}:=(k_1,\ldots,k_{\ell-1},k',\ldots,k_{m-1},-k',\ldots,k_{2n})$. It is straightforward to conclude that the above
system possesses a unique continuous solution $\rho(t,\bk)=[\rho_{\bi}(t,\bk)]$ whose components belong to $L^1(\mu)$.
The next proposition gives the convergence of even moments to the solution of (\ref{031910}).
\begin{prop}
\label{012410} 
Suppose that all the components of the tensor $[\hat M_{\bi}(d\bk)]$ are absolutely continuous with respect to $\mu$, i.e.  
$
\hat M_{\bi}(d\bk)=\rho_{\bi}(\bk)\mu(d\bk),
$
and the dispersion relation satisfies hypothesis $\om)$. 
Then, the following are true:
\begin{itemize}
\item[1)] $\hat M_{\bi}(t,d\bk)$ is absolutely continuous with respect to $\mu(d\bk)$ and
\begin{equation}
\label{012910}
\hat M_{\bi}(t,d\bk)=\rho_{\bi}(t,\bk)\mu(d\bk),\quad\forall\,\bi\in\{-1,1\}^{2n}
\end{equation}
where $\{\rho_{\bi}(t),\,t\ge0\}$ satisfy \eqref{031910a1}.
\item[2)] For any $T>0$ there exists a constant $C>0$ such that
\begin{equation}
\label{063110}
\lim_{\eps\to0+}\sum_{\bi\in\{-1,1\}^{2n}}\sup_{t\in[0,t_*]}\|\hat M_{\bi}^{(\eps)}(t)-\hat M_{\bi}(t)\|_{\rm TV}=0.
\end{equation}
\end{itemize}
\end{prop}
{\bf Proof.} 
The conclusion of part 1) follows from uniqueness of solutions of \eqref{031910} and \eqref{031910a1}, and the fact that the right hand side of \eqref{012910} defines a solution of \eqref{031910}. From \eqref{011910} and \eqref{031910} we conclude that
\begin{eqnarray}
\label{022910}
&&\|\hat M_{\bi}^{(\eps)}(t)-\hat M_{\bi}(t)\|_{\rm TV}
\le \sum_{\ell=1}^{2n}\sum_{j=\pm1}\int_0^t\left\|C_{i_\ell,j}\left(\frac{s}{\eps}\right)[\hat M_{\bi_{\ell}(j)}^{(\eps)}(s)-\hat M_{\bi_{\ell}(j)}(s)]\right\|_{\rm TV}ds\nonumber\\
&&
+\sum_{1\le \ell<m\le 2n}\sum_{j_1,j_2=\pm1}\int_0^t\left\|{\cal R}_{i_\ell,i_m}^{j_1,j_2}\left(\frac{s}{\eps}\right)[\hat M_{\bi_{\ell,m}(j_1,j_2)}^{(\eps)}(s)-\hat M_{\bi_{\ell,m}(j_1,j_2)}(s)]\right\|_{\rm TV}ds\nonumber\\
&&
+\sum_{\ell=1}^{2n}\sum_{j=\pm1}\left|\int_0^t\int_{\bbT^{2n}}E_{i_\ell,j}\left(\frac{s}{\eps},k_{\ell}\right)\rho_{\bi_{\ell}(j)}(s,\bk)ds\mu(d\bk) \right|\nonumber\\
&&
+\sum_{1\le \ell<m\le 2n}\sum_{j_1,j_2=\pm1}\left|\int_0^t\int_{\bbT^{2n+1}}\tilde{\cal R}_{i_\ell,i_m}^{j_1,j_2}\left(\frac{s}{\eps},\bk,k'\right)\rho_{\bi_{\ell,m}(j_1,j_2)}(s,\bk)ds\mu(d\bk) dk' \vphantom{\int_0^t}\right|.\nonumber
\end{eqnarray}
The matrix ${\bf E}(t,k)=[E_{p,q}(t,k)]$, $p,q=\pm1$ is given by
\begin{equation}
\label{042910}
{\bf E}(t,k):={\bf C}(t,k)-(\hat\beta(k)/4){\bf I}_2,
\end{equation} where ${\bf I}_2$ is the $2\times 2$ identity matrix. 
In addition,
$$
\tilde{\cal R}_{i_\ell,i_m}^{j_1,j_2}\left(\frac{s}{\eps},\bk,k'\right):=
{\cal R}_{i_\ell,i_m}^{j_1,j_2}\left(\frac{s}{\eps},k_\ell,k_m,k'\right)-\delta_{i_\ell}^{-i_m}
\delta_{j_1}^{-j_2}r^2(k_\ell,k_{\ell}-k') 1_{D_{\ell,m}}(\bk).
$$
Denote the terms appearing on the right hand side of \eqref{022910} by $I(t)$, $I\!I(t)$, $I\!I\!I(t)$ and $I\!V(t)$ respectively. It is easy to see that
\begin{equation}
\label{0312910}
I(t)+I\!I(t)\le C\int_0^t\sup_{\bi\in\{-1,1\}^{2n}}\left\|\hat M_{\bi}^{(\eps)}(s)-\hat M_{\bi}(s)\right\|_{\rm TV}ds
\end{equation}
for some constant $C>0$. To estimate the term $I\!I\!I$ we need to bound terms of the form 
\begin{eqnarray*}
&&
\left|\int_0^t\int_{\bbT^{2n}}\hat\beta(k_\ell)\exp\left\{2i\om(k_\ell)\frac{s}{\eps}\right\}\rho_{\bi}(s,\bk)ds\mu(d\bk)\right|
\end{eqnarray*}
for some $\ell$ and $\bi$. Integrating by parts we obtain that the expression above can be bounded from above by
\begin{eqnarray*}
&&
\eps\left|\int_{\bbT^{2n}}\frac{\hat\beta(k_\ell)}{2i\om(k_\ell)}\left[\exp\left\{2i\om(k_\ell)\frac{t}{\eps}\right\}-1\right]\rho_{\bi}(t,\bk)1_{D_{\ell,m}}(\bk)\mu(d\bk)\right|
\\
&&
+\eps\left|\int_0^t\int_{\bbT^{2n}}\frac{\hat\beta(k_\ell)}{2i\om(k_\ell)}\left[\exp\left\{2i\om(k_\ell)\frac{t}{\eps}\right\}-1\right]\frac{d}{ds}\rho_{\bi}(s,\bk)1_{D_{\ell,m}}(\bk)ds\mu(d\bk)\right|.
\end{eqnarray*}
The first term can be easily estimated by
$C\eps$, due to the fact that $\sup_{k\in\bbT}\hat\beta(k)\om^{-1}(k)<+\infty$. To estimate the second term, 
we use equation \eqref{031910a1}. As a result,, we conclude that for any $t_*>0$ we can find a constant $C(t_*)>0$ such that
\begin{equation}
\label{0512910}
\sup_{t\in[0,t_*]}I\!I\!I(t)\le C(t_*)\eps.
\end{equation}
Finally we show that
\begin{equation}
\label{102910}
\lim_{\eps\to0+}\sup_{t\in[0,t_*]}I\!V(t)=0.
\end{equation}
It implies the conclusion of  part 2) of the proposition, via an application of the Gronwall's inequality.

We write $I\!V(t)=I\!V_1(t)+I\!V_2(t)$, where the terms $I\!V_i(t)$, $i=1,2$ correspond to the integration over $D_{\ell,m}$ and its complement. In the latter case, 
we have to deal with terms of the form
\begin{eqnarray*}
&&
\left|\int_0^t\int_{\bbT^{2n+1}}1_{[k_\ell\not=-k_m]}r(k_\ell,k')r(k_m,-k')\rho_{\bi}(s,\bk)\right.\\
&&
\left.\times\prod_{j=1}^2\exp\left\{i\si_1^{(j)}[\om(k_\ell^{(j)})+\si_2^{(j)}\om(k_\ell^{(j)}+(-1)^jk')]\frac{s}{\eps}\right\}ds\mu(d\bk) dk' \vphantom{\int_0^t}\right|
\end{eqnarray*}
for some $\bi\in\{-1,1\}^{2n}$, $\si_p^{(j)}\in\{-1,1\}$. Here $k_\ell^{(1)}=k_\ell$ and $k_\ell^{(2)}=k_m$. 
Using integration by parts over the $s$ variable we can estimate the supremum of the above expression 
over $t\in[0,t_*]$ by the sum of
\begin{eqnarray}
\label{013110}
&&
I_\eps:=\int_{\bbT^{2n+1}}\mu(d\bk) dk'1_{[k_\ell\not=-k_m]}|r(k_\ell,k')r(k_m,-k')|\sup_{t\in[0,t_*]}|\rho_{\bi}(t,\bk)|\nonumber\\
&&
\times\eps\left|\sum_{j=1}^2\si_1^{(j)}[\om(k_\ell^{(j)})+\si_2^{(j)}\om(k_\ell^{(j)}+(-1)^jk')]\right|^{-1}
\\
&&
\times\sup_{t\in[0,t_*]}\prod_{j=1}^2\left|\exp\left\{i\si_1^{(j)}[\om(k_\ell^{(j)})+\
si_2^{(j)}\om(k_\ell^{(j)}+(-1)^jk')]\frac{t}{\eps}\right\}-1\right| ,\nonumber
\end{eqnarray}
and 
\begin{eqnarray}
\label{023110}
&&
J_\eps:=\int_0^Tds\left|\int_{\bbT^{2n+1}}\mu(d\bk) dk'1_{[k_\ell\not=-k_m]}r(k_\ell,k')r(k_m,-k')\frac{d}{ds}\rho_{\bi}(s,\bk)\right.\nonumber\\
&&
\times\eps\left\{\sum_{j=1}^2\si_1^{(j)}[\om(k_\ell^{(j)})+\si_2^{(j)}\om(k_\ell^{(j)}+(-1)^jk')]\right\}^{-1}
\\
&&
\left.\times\prod_{j=1}^2\left\{\exp\left\{i\si_1^{(j)}[\om(k_\ell^{(j)})
+\si_2^{(j)}\om(k_\ell^{(j)}+(-1)^jk')]\frac{s}{\eps}\right\}-1\right\}\right|. \nonumber
\end{eqnarray}
Using \eqref{031910a1} and  Gronwall's inequality, we conclude that
$$
\int_{\bbT^{2n}}\sup_{t\in[0,t_*]}|\rho_{\bi}(t,\bk)|d\bk<+\infty.
$$
Using condition $\om)$ we conclude therefore, by virtue of Lebesgue dominated convergence theorem, that
$\lim_{\eps\to0+}I_\eps=0$.
Likewise,  after substituting 
for $\rho_{\bi}'(s,\bk)$ from \eqref{031910a1}, we conclude that $\lim_{\eps\to0+}J_\eps=0$. Part 2) of the proposition 
follows then from  another application of Gronwall's inequality. Summarizing, we have shown so far that
$$
\lim_{\eps\to0+}\sup_{t\in[0,t_*]}I\!V_2(t)=0.
$$

We are left therefore with estimates of the term
\begin{eqnarray}
\label{033110}
&&
I\!V_1(t):=\sum_{1\le \ell<m\le 2n}\sum_{j_1,j_2=\pm1}\left|\int_0^t\int_{\bbT^{2n+1}}1_{D_{\ell,m}}(\bk)\right.\\
&&
\left.\times \tilde{\cal R}_{i_\ell,i_m}^{j_1,j_2}\left(\frac{s}{\eps},\bk,k'\right)\rho_{\bi_{\ell,m}(j_1,j_2)}(s,\bk)ds\mu(d\bk) dk' \vphantom{\int_0^t}\vphantom{\int_0^1}\right|.\nonumber
\end{eqnarray}
The non-vanishing terms appearing in the above sum are of the form
\[
\left|\int_0^t\int_{\bbT^{2n+1}}r^2(k_\ell,k_\ell-k')1_{D_{\ell,m}}(\bk)
\prod_{j=1}^2\exp\left\{i\si_1^{(j)}[\om(k_\ell)+\si_2^{(j)}\om(k_\ell-k')]\frac{s}{\eps}\right\}ds\mu(d\bk) dk' \vphantom{\int_0^t}\right|,
\]
with $(\si_1^{(1)},\si_2^{(1)})\not=-(\si_1^{(2)},\si_2^{(2)})$ and $\si_p^{(j)}\in\{-1,1\}$. To these terms we can apply the 
integration by parts argument as before, to conclude that
$$
\lim_{\eps\to0+}\sup_{t\in[0,t_*]}I\!V_1(t)=0.
$$
Summarizing, we have shown that \eqref{102910} holds, and the proof of part 2 of the proposition is therefore complete. \qed

\subsection{Proof of Theorem  \ref{main-thm2}} 
\label{sec5.5} 

In this section, and in this section only, 
we make use of the assumption that $\hat\psi$ is Gaussian. We show that the limiting measure for $\tilde Q_\eps$, as $\eps\to0+$, 
coincides with the law $\tilde Q$ of the process given \eqref{limit-eqt} by proving that for any $N\ge1$, $0\le t_1<\ldots <t_N$,  
any non-negative integers $\ell_j,m_j$, test functions $f_j,g_j\in H^m(\bbT)$, $j=1,\ldots,N$  we have
\begin{equation}
\label{011710}
\lim_{\eps\to0+}\bbE\left[\prod_{j=1}^N[\langle\tilde\psi^{(\eps)}(t_j),f_{j}\rangle^{\ell_j}(\langle\tilde\psi^{(\eps)}(t_j),g_{j}\rangle^*)^{m_j}]\right]
=
\bbE\left[\prod_{j=1}^N[\langle\bar\psi(t_j),f_{j}\rangle^{\ell_j}(\langle\bar\psi(t_j),g_{j}\rangle^*)^{m_j}]\right]. 
\end{equation}
To simplify the notation, we  prove \eqref{011710} only in the case $N=1$. The general case can be handled in the same manner, 
using Markov property of the process  $\{\tilde\psi^{(\eps)}(t),\,t\ge0\}$, at the expense of some additional complications 
in the notation. We recall (see Section \ref{sec2.3.2})  that the initial data $\{ \hat\psi(k),\,k\in\bbT\}$ is a  $\delta$-correlated Gaussian random field given by \eqref{053110}. Therefore, for the odd moments we have 
$$
\hat M^{(\eps)}_{\bi}(0)=
0,\,\quad \forall\,\bi\in\{-1,1\}^{2n-1},
$$
where $n\ge 1$ is an integer. By uniqueness of solutions of \eqref{011910} we conclude that
in this case $
\hat M^{(\eps)}(t)\equiv0$
for all $t\ge0$. When $\bi\in \{-1,1\}^{2n}$ we can use the conclusion  \eqref{063110} of Proposition~\ref{012410}.
Define 
$$
\bar M^{(2n)}(t):=
\left[ \bar M_{\bi}^{(2n)}(t)\right],\,\quad \bi=(i_1,\ldots,i_{2n})\in\{-1,1\}^{2n},
$$
where
$$
\bar M_{\bi}^{(2n)}(t) =\bbE\left[\bar \psi_{i_1}(t)\otimes \ldots\otimes \bar \psi_{i_{2n}}(t)\right]
$$
and $\bar \psi_{1}(t)=\bar \psi(t)$ is the solution of \eqref{limit-eqt} and $\bar \psi_{-1}(t,k)=\bar \psi^*(t,-k)$.
 The  conclusion of Theorem \ref{main-thm2} will follow 
 provided that we  show that $\bar M^{(2n)}(t)$, satisfies  \eqref{031910}.  Note that for $n=1$ we obtain that
 $$
\bar M_{i_1,i_2}^{(2)}(t,d\bk) =\delta_{i_1,-i_2}\bar{\cal E}(t,k_1)\delta(k_1+k_2)dk_1dk_2.
$$
From \eqref{limit-eqt} and It\^o formula we conclude that
\begin{eqnarray}
&&
 \frac{d}{dt}\bar M_{\bi}^{(2n)}(t)=-\frac{1}{4}\left(\sum_{\ell=1}^{2n}\hat\beta\left(k_\ell\right)\right)\bar M_{\bi}^{(2n)}(t)
 -\sum_{1\le \ell<m\le 2n}{\cal R}(t,k_{\ell}) \bar M_{\bi_{\ell,m}}^{(2n-2)}(t)\otimes_{\ell,m}\Delta ,\nonumber
\label{031910a}\\
&&
\bar M(0)=\hat M.
\end{eqnarray} 
Here $\bar M_{\bi_{\ell,m}}^{(2n-2)}(t)$ is the $2n-2$-nd order moment obtained from $\bar M_{\bi}^{(2n)}(t)$ by omitting $\bar \psi_{i_\ell}(t)$ and $\bar \psi_{i_m}(t)$ and for any measure $\nu$ on $\bbT^{2n-2}$, $1\le \ell<m\le 2n$ we denote by $\nu\otimes_{\ell,m}\Delta$  a measure on $\bbT^{2n}$ given by
\begin{eqnarray*}
&&
\int_{\bbT^{2n}}fd(\nu\otimes_{\ell,m}\Delta)
=\int_{\bbT^{2n-2}}d\bk\int_{\bbT}dk f(k_1,\ldots,k_{\ell-1},k,\ldots,k_{m-1},-k,\ldots,k_{2n-2})
\end{eqnarray*}
for all $f\in C(\bbT^{2n})$. Since
\begin{eqnarray*}
&&
{\cal R}(t,k_{\ell})=\int_{\bbT}R(k_\ell,k')\bar{\cal E}(t,k')dk'=\int_{\bbT}[r^2(k_\ell,k_\ell-k')+r^2(k_\ell,k_\ell+k')]\bar{\cal E}(t,k')dk'
\\
&&
=\sum_{j=\pm1}\int_{\bbT^2}r^2(k_\ell,k_\ell-k')\bbE\left[\bar\psi_j(t,k')\otimes \bar\psi_{-j}(t,k'')\right]dk'dk''
\end{eqnarray*}
and 
$(\bar \psi_{i_1}(t),\ldots, \bar \psi_{i_{2n}}(t))$ is jointly Gaussian, 
we infer that  the last term on the right hand side of the first equation in \eqref{031910a}  equals  the last term on the right hand side of the first equation of \eqref{031910}. Thus the conclusion of 
Theorem~\ref{main-thm2} has been shown.

\appendix

\section{Proof of Proposition \ref{prop010910}}

\label{appA}

To prove the proposition we verify that  for any $T>0$
 \begin{equation}
\label{030910}
 t\mapsto  {\cal A}[t,\cdot]\quad \mbox{ is  Lipschitz on }H^{m}(\bbT),
 \end{equation}
  uniformly in $t\in[0,T]$ and
 $\tilde Q[t,g] :L^2(\bbT)\to H^m(\bbT)$, given by
\eqref{022808} is Hilbert-Schmidt  for any $g\in H^m(\bbT)$ and its respective Hilbert-Schmidt norm satisfies
\begin{equation}
\label{020910a}
\sup_{t\in[0,T]}\| \tilde Q[t,g_1]-\tilde Q[t,g_2]\|_{(HS)_m}\le C\|g_1-g_2\|_{H^m(\bbT)},\quad\forall\,g_1,g_2\in  H^m(\bbT)
\end{equation}
for some $C>0$. 
The conclusion of the lemma then follows from  \cite{DZ}, Theorem~7.4, p. 186.

 Since $\beta_x\not=0$ only for $|x|\le 2$, see \eqref{040910}, to prove \eqref{030910} it suffices only to show that there exists $C>0$ such that
 \begin{equation}
 \label{060910}
 \sup_{t\in[0,T]}\|f(t)\|_{H^m(\bbT)}\le C\|f\|_{H^m(\bbT)},\quad\forall\,f\in H^m(\bbT),
 \end{equation}
 with $f(t):=\exp\left\{2i\om(k)t\right\}f(k)$.
 Dispersion relation $\om(\cdot)$ given by \eqref{om} is bounded with its all
derivatives on $\bbT\setminus\{0\}$. In addition $\om'(0-)$ and $\om'(0+)$ exist. Therefore
\begin{equation}
\label{050910}
\om_*:=\sup_{t\in[0,T],\,x\in\bbZ}(1+x^2)\left|\gamma_x(t)\right|<+\infty,
\end{equation}
where
$$
\gamma_x(t):=\int_{\bbT}\exp\left\{2i\om(k)t\right\}e_x^*(k)dk.
$$
Note that
\begin{equation}
\label{031010}
1+y^2\le \sup_x\frac{1+x^2}{1+(x-y)^2}\le 2(1+y^2).
\end{equation}
Assume first that $m\ge0$. We can write then
\begin{eqnarray}
\label{011110}
&&
\|f(t)\|_{H^m(\bbT)}^2=\sum_{x\in\bbZ}(1+x^2)^{m}\left|\check f_x(t)\right|^2
=\sum_{x\in\bbZ}(1+x^2)^{m}\left|\sum_{y\in\bbZ}\check f_{x-y}\gamma_y(t)\right|^2\\
&&
=\sum_{x\in\bbZ}\left|\sum_{y\in\bbZ}(1+(x-y)^2)^{m/2}\check f_{x-y}\frac{(1+x^2)^{m/2}\gamma_y(t)}{(1+(x-y)^2)^{m/2}}\right|^2.\nonumber
\end{eqnarray}
Using \eqref{050910}  together with   \eqref{031010} we can
%
we can estimate the utmost right hand side of \eqref{011110}
 by
$$
 2\om_*^2
\sum_{x\in\bbZ}\left\{\sum_{y\in\bbZ}(1+(x-y)^2)^{m/2}|\check f_{x-y}|(1+y^2)^{m/2-1}\right\}^2.
$$
Using Young's inequality $\|f*g\|_{\ell^r} \le
\|f\|_{\ell^p}\|g\|_{\ell^q}$, where $1+r^{-1} = p^{-1} + q^{-1}$,
(with $r=p=2$, $q=1$) 
we can bound this expression by 
$$
C\left\{\sum_{x\in\bbZ}(1+x^2)^{m}|\check f_{x}|^2\right\}
\left\{\sum_{y\in\bbZ}(1+y^2)^{m/2-1}\right\}^2
$$
for some constant $C>0$. Summarizing we have shown that
$$
\|f(t)\|_{H^m(\bbT)}^2\le  2\om_*^2
\|f\|_{H^m(\bbT)}^2\left\{\sum_y(1+y^2)^{m/2-1}\right\}^2,
$$
which proves \eqref{060910}, provided $0\le m<1$.

If, on the other hand, $m<0$ we  can write
\begin{eqnarray*}
&&
\sum_{x\in\bbZ}(1+x^2)^{m}\left|\check f_x(t)\right|^2
\le \om_*^2
\sum_{x\in\bbZ}(1+x^2)^{m}\left[\sum_{y\in\bbZ}(1+y^2)^{-1}|\check f_{x-y}|\right]^2.
\end{eqnarray*}
By Cauchy-Schwartz inequality for any $\gamma>1/2$ the right hand side can be estimated by
\begin{equation}
\label{011310-2012}
 \om_*^2
\sum_{x\in\bbZ}(1+x^2)^{m}\left[\sum_{y\in\bbZ}(1+y^2)^{-\gamma}\right]\left[\sum_{y\in\bbZ}(1+(x-y)^2)^{-(2-\gamma)}|\check f_{y}|^2\right].
\end{equation}
We use the following elementary inequality: for any $\ka>1/2$ there exists a constant $C>0$ such that
\begin{equation}
\label{021110}
\sum_{x\in\bbZ}(1+x^2)^{m}(1+(x-y)^2)^{-\kappa}\le C(1+y^2)^{m\vee (-\ka)},\quad \forall y\in\bbZ.
\end{equation}
%
%
%
%
%
%
%
%
%
%
%
%
%
%
Let $\ka:=2-\ga$ and $\ga\in(1/2,3/2)$. We conclude from the above estimate that
the expression in \eqref{011310-2012} is less than, or equal to
$
C\|f\|_{H^m(\bbT)}^2,
$ provided that $2+m>\ga$, which is possible as long as $m>-3/2$. 

%
%
%
%
%
%
%
%
%
%
To show \eqref{020910a} it suffices to prove that for any functions $\phi_1,\phi_2$ that are finite combinations of the vectors from the base $(e_x)$  and $T>0$ there exists a constant $C>0$ such that
\begin{eqnarray}
\label{011010}
&&
\sup_{t\in[0,T]}\sum_{x,y}(1+y^2)^m\left|\int_{\bbT^2}\phi_1(k)\phi_2(k-k')g(k-k')e_x(k') e_y^*(k) 
\exp\left\{i[\om(k)-\si\om(k-k')]t\right\}dk dk'\right|^2\nonumber\\
&&
\le C\|g\|_{H^m(\bbT)}^2,\quad \forall\, g\in H^m(\bbT),\, \sigma = \pm 1.
\end{eqnarray}
The expression on the left hand side of \eqref{011010} can be rewritten in the form
\begin{equation}
\label{021010}
\sum_{x,y}(1+y^2)^m\left|\psi_{y-x}^{(1)}(t)\psi_{x}^{(2)}(t)\right|^2,
\end{equation}
where
$$
\psi_x^{(1)}(t):=\int_{\bbT}\phi_1(k) e_x^*(k) 
\exp\left\{i\om(k)t\right\}dk
$$
and
$$
\psi_x^{(2)}(t):=\int_{\bbT}\phi_2(k)g(k)e_x^*(k) 
\exp\left\{-i\si\om(k)t\right\}dk. 
$$
As a consequence of \eqref{060910}  for any $T>0$ there exists $C>0$ such that
$$
\sup_{t\in[0,T]}\sum_{x}(1+x^2)^m\left|\psi_{x}^{(2)}(t)\right|^2\le C\|g\|_{H^m(\bbT)}^2,\quad \forall\,g\in H^m(\bbT).
$$
We also have 
$
\sup_{t\in[0,T]}(1+x^2)|\psi_x^{(1)}(t)|<+\infty.
$
The expression in \eqref{021010} can be rewritten as
\begin{eqnarray}
\label{051010}
&&
\sum_{x,z}(1+(z+x)^2)^m\left|\psi_{z}^{(1)}(t)\psi_{x}^{(2)}(t)\right|^2\\
&&
=\sum_{x,z}\frac{(1+(z+x)^2)^m}{(1+x^2)^m}(1+x^2)^m\left|\psi_{z}^{(1)}(t)\psi_{x}^{(2)}(t)\right|^2.\nonumber
\end{eqnarray}
Suppose that $m\ge0$ then the right hand side  can be estimated by
\begin{eqnarray*}
&&
2^m\sum_{x,z}(1+z^2)^m(1+x^2)^m\left|\psi_{z}^{(1)}(t)\psi_{x}^{(2)}(t)\right|^2\\
&&
=
2^m\left(\sum_{z}(1+z^2)^m\left|\psi_{z}^{(1)}(t)\right|^2\right)\left(\sum_{x}(1+x^2)^m\left|\psi_{x}^{(2)}(t)\right|^2\right)\\
&&
\le C\left(\sum_{z}(1+z^2)^{m-2}\right)\|g\|_{H^m(\bbT)}^2,
\end{eqnarray*}
which proves \eqref{020910a}, provided that $0\le m<3/2$.

If, on the other hand, $m<0$ the left hand side of \eqref{051010} can be estimated by
\begin{eqnarray}
\label{051010a}
&&
C\sum_{x,z}(1+(z+x)^2)^m(1+z^2)^{-2}\left|\psi_{x}^{(2)}(t)\right|^2\\
&&
\le C_1\sum_{x}(1+x^2)^{m\vee (-2)}\left|\psi_{x}^{(2)}(t)\right|^2\le C_2\|g\|_{H^m(\bbT)}^2,\nonumber
\end{eqnarray}
provided that $m>-2$.

%
%
%
%

\end{document}